\documentclass[11pt]{article}
\usepackage{comment}

\usepackage{amsmath}
\usepackage{amsthm}
\usepackage{listings}
\usepackage{amssymb}
\usepackage{algorithm,algorithmic}
\usepackage{dblfloatfix}
\usepackage{empheq}
\usepackage[margin=1in]{geometry}

\usepackage{hyperref}
\usepackage[capitalize]{cleveref}

\usepackage{thm-restate}

\newtheorem{theorem}{Theorem}
\newtheorem{corollary}[theorem]{Corollary}

\newtheorem{definition}[theorem]{Definition}
\newtheorem{proposition}[theorem]{Proposition}

\newtheorem{remark}[theorem]{Remark}

\usepackage[]{color-edits}
\addauthor{UF}{red}
\newcommand{\ufc}[1]{{\UFcomment{#1}}}
\newcommand{\ufe}[1]{{\UFedit{#1}}}

\addauthor{AN}{blue}

\newcommand{\ane}[1]{{\ANedit{#1}}}

\newcommand{\OLD}[1]{}

\newcommand{\items}{{\cal{M}}}
\newcommand{\agents}{{\cal{N}}}

\begin{document}

\title{Improved maximin fair allocation of indivisible items to three agents}
\author{Uriel Feige\thanks{uriel.feige@weizmann.ac.il} \; and Alexey Norkin\thanks{alexnorkin97@gmail.com} \\ Weizmann Institute}

\maketitle

\begin{abstract}
We consider the problem of approximate \emph{maximin share} (MMS) allocation of indivisible items among three agents with additive valuation functions. For goods, we show that an $\frac{11}{12}$-MMS allocation always exists, improving over the previously known bound of $\frac{8}{9}$. Moreover, in our allocation, we can prespecify an agent that is to receive her full \emph{proportional share} (PS); we also present examples showing that for such allocations the ratio of $\frac{11}{12}$ is best possible. For chores, we show that a $\frac{19}{18}$-MMS allocation always exists.
Also in this case, we can prespecify an agent that is to receive no more than her PS, and we  present examples showing that for such allocations the ratio of $\frac{19}{18}$ is best possible.
    
\end{abstract}

\section{Introduction}

We consider allocation of $m$ indivisible items to $n$ agents with additive valuations. An item that is valued non-negatively by an agent is referred to as a \emph{good}.  An item that is valued non-positively by an agent is referred to as a \emph{chore}. In our setting there are no monetary transfers. We wish our allocation to meet a certain fairness criterion. 

For divisible items, a well studied fairness criterion is to provide every agent a bundle of value at least her \emph{proportional share} (PS), namely, at least a $\frac{1}{n}$ of the total value of all items, with respect to her valuation function. However, for indivisible items, there are instances in which this criterion cannot be met, not even approximately (e.g., when there are fewer goods than agents). 

A more suitable fairness criterion for indivisible items is the \emph{maximin share} (MMS), namely, the maximum over all partitions into $n$ bundles, of the the smallest value of a bundle under the agent's valuation function.  An {\em MMS-allocation} is an allocation in which if every agent receives her MMS.

Of course, if all agents have the same valuation function, an MMS allocation exists. In fact, it suffices that all but one agent have the same valuation function. The latter includes the case $n=2$ which allows for MMS allocations (by the standard cut and choose procedure). However, perhaps surprisingly, for $n\ge 3$ there are instances that admit no MMS allocation, as was shown in \cite{PKW18}. This fact initiated the study of $\rho$-MMS allocations. For goods, a $\rho$-MMS allocation needs to give every agent a bundle of value at least a $\rho$ fraction of her MMS (and $\rho \le  1$), whereas for chores it needs to give every agent a bundle of value at most a $\rho$ fraction of her MMS (and $\rho \ge 1$). Whereas there are no tight results known yet for the value of $\rho$, it is known that for any instance of goods $\rho\ge \frac{3}{4}+\frac{1}{12n}$ (see \cite{GT20}), and for any instance of chores $\rho\le \frac{11}{9}$ (\cite{HL21}). For three agents the tighter results are known. In particular, for goods it is known that $\rho \ge  \frac{8}{9}$(\cite{GM19}) and $\rho \le  \frac{39}{40}$ (\cite{FST21}).

In our work, we improve the known values of $\rho$ for the case of $n=3$, for goods and for chores. Moreover, we do so while ensuring that a prespecified agent gets at least her PS. We refer to such allocations $\rho$-P$^+$MM. (In this notation, P stands for PS, M stands for MMS, a sequence PMM ordered as such designates that the agent who gets her PS is pre-specified, and a superscript $^+$ designates that the agent receives her full share instead of approximate one.)

For goods we prove that $\rho \ge \frac{11}{12}$, and for chores we prove that $\rho \le \frac{19}{18}$. Our results are tight for $\rho$-P$^+$MM, and remain tight even if we relax $\rho$-P$^+$MM to $\rho$-PMM (for which it suffices to give the P agent a $\rho$-fraction of her proportional share).  That is, we provide an instance for goods (and an instance for chores) that admits no $\rho$-PMM allocation with $\rho$ better than $\frac{11}{12}$ (better than $\frac{19}{18}$ for chores).

Our proofs for the approximation ratios are computer assisted. Using such analysis was inspired by earlier work \cite{FST21}. We provide the full code so that readers can verify its correctness and run it independently to verify also that it gives the results we claim. Negative examples, on the other side, are given explicitly, and can be verified without the aid of a computer.

{In passing, we revisit the earlier works of \cite{AMNS17,GM19} that provided for three agents a $\frac{7}{8}$ and $\frac{8}{9}$-MMS allocation for goods, respectively.} The allocation algorithm presented in these works in fact ensures that one pre-specified agent gets her full PS, one pre-specified agent gets her full MMS, and only the remaining agent settles for $\rho$-MMS. We show that the allocation algorithm in these works leads to a value of $\rho$ better than shown in these works, namely, $\frac{9}{10}$, and also provide an example showing that the ratio provided by their algorithm is not better than $\frac{9}{10}$.

Finally, we also observe that for every fixed $n$, the best value of $\rho$ for which $\rho$-MMS allocations exist is a solution to a mixed integer/linear program (MILP) of finite size, and hence in principle can be computed exactly. However, even for $n=3$ this MILP appears to be too large to be solved in practice. 

\subsection{Related Work}
{
We examine the problem of fair allocation of a set of items among agents with additive valuation functions. We focus on fair division of indivisible items, though note that there is also work (such as \cite{S49} or \cite{V74}) on  allocation of divisible items with various notions of fairness. 

Different concepts of fairness were introduced over the course of time. The notion of envy-freeness (EF) was introduced by \cite{GS58}. For divisible items, EF allocations do not always exist. In fact, determining whether a given instance allows for EF allocations or not is NP-complete~\cite{LMMS04}. Thus, certain relaxations of EF were proposed. One of them is envy free up to one item (EF1)~\cite{B11}. There exist simple polynomial algorithms that produce EF1 allocations~\cite{LMMS04}. A stricter notion of fairness, envy free up to any good (EFX), was introduced in \cite{CKMPSW19b}. Whereas the existence of EFX allocations in general is an open problem, they were shown to exist for three agents with additive valuations (see \cite{CKMS21}).

In this work we focus on a share based approach to the concept of fairness, and specifically consider maximin share fairness (MMS), introduced in~\cite{B11}. Computing the MMS value of an agent is an NP-hard problem; however, there are polynomial time algorithm that approximate the value of the MMS within any (fixed) desired level of precision (\cite{W97}). MMS allocation exist for any instance with two agents and additive valuations, but for three or more agents, there are instances (with additive valuation functions) for which no MMS allocation exists~\cite{PKW18}. 

Several algorithms were designed that produce approximate MMS allocations for goods, that is, $\rho$-MMS-allocations, with $\rho\in(0,1)$. 
The currently highest value of $\rho$ achieved for a general number $n$ of agents is $\frac{3}{4}+\frac{1}{12n}$~\cite{GT20}. For $n=3$, a ratio of $\rho=\frac{7}{8}$ was shown in~\cite{AMNS17}, and improved analysis of the underlying algorithm improved the ratio to $\rho=\frac{8}{9}$~\cite{GM19}. In terms of upper bounds, an instance with three agents with $\rho \le  \frac{39}{40}$ was designed in~\cite{FST21}.   

In this work we also consider approximate MMS-allocation for the case of additive valuations over chores. Instances in which an MMS allocations does not exist were shown in~\cite{ARSW17}. The current best approximation (that holds for any number of agents) is $\frac{11}{9}$ for any instance~\cite{HL21}.   In terms of lower bounds, an instance with three agents with $\rho\ge \frac{44}{43}$ was designed in \cite{FST21}.

In this paper we consider MMS allocations for additive valuation functions. For some previous work on MMS allocations for non-additive valuations, see~\cite{GHSSY18}.

}

\subsection{Definitions, Notation and Preliminaries}

In this section we introduce basic definitions and notation. $\items = \{e_1,\ldots,e_m\}$ denotes the set of items and $\agents = \{1,\ldots,n\}$ denotes the set of agents. Every agent $i$ has an additive valuation function $v^i:2^M\rightarrow\mathbb{R}$. Namely, for every bundle $S \subset \items$, $v^i(S) = \sum_{e \in S} v^i(e)$.  If all the valuation functions are non-negative for the given sets of agents and items, we call such items $goods$. Similarly, if all the valuations functions are non-positive, we call such items $chores$. {The ordered sequence $I=(\items, \agents, (v^i)^{i\in\agents})$ is called an instance. When dealing with chores it is useful to define $-I=(\items, \agents, (-v^i)^{i\in\agents})$.}

$P_n(\items)$ denotes the set of all possible $n$-partitions of $\items$. An allocation $A=(A_1, \ldots, A_n)$ is an $n$-partition of $\items$, where every agent $1\le  i \le  n$ receives the bundle $A_i$. For goods we consider the maximin share:

\begin{definition}
The maximin share of agent i, which we denote as $MMS^i$, is the maximum over all possible $n$-partitions of $\items$ of the minimum value under $v^i$ of a bundle in each partition. That is,
$$MMS^i:=\max_{(B_1,\ldots,B_n)\in P_n(\items)} \min_{j}v^i (B_j).$$
\end{definition}

For chores, when translating the instance $I$ to the instance $-I$, we consider the minimax share:

\begin{definition}
The minimax share of agent i, which we denote as $mmS^i$, is
$$mmS^i:=\min_{(B_1,\ldots,B_n)\in P_n(\items)} \max_{j}v^i (B_j).$$
\end{definition}

We shall consider allocations that give each agent a bundle that approximates the agent's share value within a multiplicative factor of $\rho$.

\begin{definition}
For $\rho > 0$, an allocation $A\in P_n(\items)$ is a $\rho$-maximin allocation iff for every agent $i$
$$v^i(A_i)\ge  \rho \cdot MMS^i.$$
\end{definition}

\begin{definition}
For $\rho > 0$, an allocation $A\in P_n(\items)$ is a $\rho$-minimax allocation iff for every agent $i$
$$v^i(A_i)\le  \rho \cdot mmS^i.$$
\end{definition}

From those definitions follows an obvious relation between the maximin share and the minimax share:

\begin{proposition}
\label{pro: minimax equivalence}
An allocation $A=(A_1,\ldots,A_n)$ is the solution of the $\rho$-maximin share problem for instance $I$ iff $A$ is the solution of the $\rho$-minimax share problem for instance $-I$.
\end{proposition}

Finally, we define a proportional share:

\begin{definition}
The proportional share of agent $i$, which we denote as $PS^i,$ is
$$PS^i:=\frac{\sum\limits^{m}_{j=1} v^i (e_j)}{n}.$$
\end{definition}

Using those definitions, we immediately obtain an inequality between the aforementioned shares:

\begin{proposition}
For any given instance $I,$ the following relation holds for every agent $i$:
$$MMS^i \le  PS^i \le  mmS^i.$$
\end{proposition}

We shall design allocation algorithms that provide some agents their approximate maximin (or minimax) shares, and provide other agents their approximate (or exact) proportional share. For this purpose, we introduce some notation. We shall first present the notation for the case of goods (where it involves the maximin share), and later extend it to the case of chores (where it involves the minimax share). 

For an agent that needs to get the $MMS$ we use the letter `$M$', and for an agent that needs to get the $PS$ we use the letter `$P$'. So, an $MMP$ allocation  means that two agents receive their maximin shares and the other agent receives her proportional share. However, if we wish to fix in advance the agent that needs to receive her proportional share, we denote such an allocation $PMM$.

When we refer to approximate allocations we add $\rho$ at the beginning, extending the notation $\rho$-$MMS$. 
When we have a $P$ agent then $\rho$-$MMP$ (or $\rho$-$PMM$, if the $P$ agent is fixed in advance) denotes an allocation in which the $M$ agents get at least a $\rho$ fraction of their MMS, whereas the $P$ agent gets at least a $\rho$ fraction of her PS. However, we shall also consider allocations in which the $\rho$ approximation refers only to the $M$ agents,  whereas the $P$ agent gets her full $PS$. We denote such allocations by $\rho$-$MMP^+$ (or $\rho$-$P^+ MM$, if the $P$ agent is fixed in advance).

Clearly, every $\rho$-$MMP^+$ allocation is also a $\rho$-$MMP$ allocation. As the PS is at least as large as the MMS, every $\rho$-$MMP$ allocation is also a $\rho$-$MMS$ allocation. Interestingly, a converse relation also holds: every $\rho$-$MMS$ allocation can be transformed into a $\rho$-$MMP^+$ allocation. 
{
\begin{proposition}
For any instance $I$ of $n$ agents (with additive valuations) and $m$ items, every $\rho$-$MMS$ allocation $A$ can be transformed into a new $\rho$-$MMS$ allocation  $A^*$ in which at least one agent gets her proportional share.
\end{proposition}

\begin{proof}
Given an allocation $A = (A_1, \ldots, A_n)$, we say that agent $i$ {\em envies} agent $j$ if $v^i(A_i)<v^i(A_j)$. We say that agent $i$ is {\em envy-free} if she does not envy any other agent.
Observe that an envy-free agent gets her proportional share. Hence to prove the proposition it suffices to consider those $\rho$-$MMS$ allocations that do not have an envy-free agent.

An allocation $A$ is said to {\em Pareto dominate} an allocation $A'$, if every agent gets in $A$ value at least as high as in $A'$, and at least one agent gets strictly higher value. Pareto domination induces a partial order over allocations, and the maximal elements in this partial order are the Pareto optimal (PO) allocations.

Given a $\rho$-$MMS$ allocation $A$ in which there is no envy-free agent, consider any PO allocation $A'$ that Pareto dominates $A$. Clearly, $A'$ is a  $\rho$-$MMS$ allocation. We claim that $A'$ has an envy-free agent. The claim implies that $A'$ has an agent that receives her proportional share, as desired.

To prove the claim, consider the directed {\em envy graph} associated with the PO allocation $A'$. The vertices are the agents, and for every pair of agents $(i,j)$, we include in the graph the directed edge $(i,j)$ if $i$ envies $j$. If there are no envy-free agents, every vertex has an outgoing edge, and the envy graph must contain at least one directed cycle. Within the cycle, we can rotate the bundles among the agents (if $(i,j)$ is a directed edge in the cycle, then $i$ gets the bundle of $j$) so that every agent along the cycle gets a bundle that she values more than her original bundle. This contradicts the assumption that $A'$ is PO.
\end{proof}

We shall use $m$ notation instead of $M$ when considering minimax shares {(for chores)}. For example, $\rho$-$P^+mm$ means that one agent (fixed in advance) needs to get at most her full PS, and the other two agents get at most $\rho$ times their minimax shares.
}

\section{Allocation Algorithms}
\label{sec:algorithms}

For simplicity of the presentation, we focus first on allocation of goods, and describe extensions to chores only later. Section~\ref{sec:MILP} contains the proofs for most of the theorems and propositions that appear in this section.

In our work we consider $n=3$, and analyse the value of $\rho$ offered by $\rho$-$MMS$ allocations and by $\rho$-$PMM$ allocations, where the allocations are produced by fairly straightforward algorithms. Before focusing on $n=3$, we consider the case of general $n$. 

Given an allocation instance $I=(\items, \agents, (v^i)^{i\in\agents})$, all our allocation algorithms first compute an optimal MMS partition for every agent, and in the process obtain the value $MMS^i$ for every agent $i$. This part of the allocation algorithm need not run in polynomial time.

\subsection{Exhaustive search}

The baseline algorithm against which we compare other algorithms is {\em exhaustive search}.  For an allocation $A = (A_1, \ldots, A_n)$, define $\rho_A$ to be $\min_{i \in \agents}[\frac{v^i(A_i)}{MMS^i}]$. The algorithm tries all possible allocations (partitions in $P_n(\items)$), and selects the allocation $A$ with highest value of $\rho_A$ (breaking ties arbitrarily). For every instance, the exhaustive search algorithm is optimal in terms of the value of $\rho$ that it guarantees. However, from our point of view, it has two drawbacks. One drawback is that even if the MMS partitions are given, the algorithm's running time is exponential in $m$, as it explicitly goes over all $n^m$ possible allocations. The other drawback is that it is not clear how to analyse what value of $\rho$ the algorithm guarantees.

\subsection{Atomic exhaustive search}
\label{sec:27}

Our next algorithm is referred to as {\em atomic exhaustive search}. In its first phase, the algorithm uses the MMS partitions so as to partition the items into $n^n$ disjoint {\em atomic bundles} (where some of the bundles may be empty). Each atomic bundle is the intersection of $n$ bundles, one bundle from the MMS partition of each of the $n$ agents. We say that an allocation respects the atomic bundles (or for brevity, that the allocation is atomic) if every atomic bundle is contained in a single bundle of the allocation. That is, an atomic bundle cannot be broken by the allocation (in this respect the atomic bundle is ``atomic"). In its second stage, the algorithm tries all possible atomic allocations, and selects the atomic allocation $A$ with highest value of $\rho_A$ (breaking ties arbitrarily).

Compared to the exhaustive search algorithm, the atomic exhaustive search algorithm tries out fewer allocations (unless every atomic bundle is either empty or contains one item). Hence, on every given instance, the value of $\rho$ for the latter algorithm might not be as good as that for the former algorithm. However, as implied by the following proposition, for every given $n$, the worst possible value of $\rho$ over all instances with $n$ agents is the same for both algorithms. 

\begin{proposition}
\label{pro:AES}
Let $\rho < 1$ be such that there is an allocation instance $I$ with $n$ agents and additive valuations on which atomic exhaustive search produces an allocation in which there is an agent that does not get more than a $\rho$ fraction of her MMS. Then there is an allocation instance $I'$ with $n$ agents and additive valuations in which in every allocation some agent does not get more than a $\rho$ fraction of her MMS.
\end{proposition}

\begin{proof}
Given $I$, generate $I'$ by replacing each atomic bundle of $I$ by a single item in $I'$, where for each agent, the value of the item in $I'$ equals the value of the corresponding atomic bundle in $I$. For every agent, her MMS value in $I'$ is the same as her MMS value in $I$. Every allocation in $I'$ naturally corresponds to an atomic allocation in $I$, and each agent receives the same value in both allocations.
\end{proof}


Atomic exhaustive search has two notable advantages over exhaustive search. One advantage is that given the MMS partitions, its running time is polynomial (in fact, linear) in $m$ (though not polynomial in $n$). Specifically, its first stage takes time $O(n^n \cdot m)$, whereas its second stage is independent of $m$ and takes time $O(n^{n^n})$. For constant $n$, the running time is $O(m)$, though the $O$ notation hides a constant that depends on $n$ in a rather bad way. The other advantage is that there is a plausible approach for analysing the value of $\rho$ guaranteed by atomic exhaustive search.

\begin{theorem}
\label{thm:AES}
For every given $n$, the value of $\rho$ guaranteed by atomic exhaustive search can be computed by solving a mixed integer/linear program (MILP) of finite size (that depends on $n$). This MILP has integer coefficients and is feasible and bounded, and hence this worst possible value of $\rho$ is rational.
\end{theorem}

Let $\rho_{n,m}$ denote the largest value of $\rho$ such that every instance with $n$ agents and $m$ items (with additive non-negative valuations) has a $\rho$-MMS allocation. (It will follow shortly that this maximum exists, and is not just a supremum.) Let $\rho_n:=\min_m \rho_{n,m}$. (Likewise, it will follow that a minimum exists, not just an infimum.)

\begin{corollary}
\label{cor:AES}
For every $n$, the value of $\rho_n$ (as defined above) is rational.
\end{corollary}

\begin{proof}
Proposition~\ref{pro:AES} implies that $\rho_n$ is the same as the value of $\rho$ guaranteed by atomic exhaustive search for all instances with $n$ agents (and additive valuations over goods). Theorem~\ref{thm:AES} implies that this latter value is rational.
\end{proof}

Previously, MILPs were used in~\cite{FST21} in order to determine the optimal value of $\rho$ (for the exhaustive search algorithm) over all instances with three agents and nine items. The principles for using MILPs in the analysis of atomic exhaustive search are similar to those used in~\cite{FST21}. However, the MILPs of Theorem~\ref{thm:AES} are huge, and consequently, solving the associated MILP might not be feasible in practice, not even for $n=3$. (Linear programs (LPs) can be solved in polynomial time, but solving MILPs is NP-hard. Hence they can be solved in practice only if the number of integer variables is not too large. The number of integer variables in the MILP of Theorem~\ref{thm:AES} is exponential in the number of atomic bundles. For the smallest value of $n$ of interest, namely, $n=3$, the number of atomic bundles is $3^3 = 27$.)

\subsection{Coarse atomic exhaustive search}
\label{sec:9G}

From here on we focus on the case that $n=3$. To reduce the size of the MILPs that are used in the analysis of our algorithms, we introduce a new algorithm that we refer to as {\em coarse atomic exhaustive search}. It has fewer atomic bundles than the atomic exhaustive search algorithm, leading to a smaller size MILP for its analysis. Specifically, we name the three agents as $R$ (for {\em row}), $C$ (for {\em column}) and $U$. The atomic bundles are the intersections of the MMS bundles of $R$ and $C$, whereas the MMS bundles of $U$ play no role in this respect. Consequently, there are only nine atomic bundles. Arranging the atomic bundles in a three by three matrix as depicted in  Section~\ref{sec:terminology}, the bundles of the MMS partition for $R$ are the rows of this matrix, and the bundles of the MMS partition for $C$ are the columns.  (The same set of nine atomic bundles was previously used in a previous algorithm of \cite{AMNS17}. See more details in Section~\ref{sec:9P}.) The coarse atomic exhaustive search algorithm tries all possible atomic allocations (with respect to these nine atomic bundles), and selects the atomic allocation $A$ with highest value of $\rho_A$ (breaking ties arbitrarily).

Compared to the atomic exhaustive search algorithm, the coarse atomic exhaustive search algorithm tries out fewer allocations. Hence, on every given instance, the value of $\rho$ for the latter algorithm might not be as good as that for the former algorithm. Moreover, the worst possible value of $\rho$ over all instances with three agents need not be the same for both algorithms (and we conjecture that the value of $\rho$ for atomic exhaustive search is strictly better than that for coarse atomic exhaustive search). However, coarse atomic exhaustive search has the advantage that the associated MILP that determines its guarantee for $\rho$ can be solved in practice (after making some simplifications to the MILP that are shown not to affect the end result), using standard MILP solvers that are publicly available on the web (we used https://online-optimizer.appspot.com/). 

Using an MILP to analyse coarse atomic exhaustive search we show that for $n=3$ (and additive valuations over goods), there always is a $\rho$-$MMS$ allocation with $\rho \ge \frac{11}{12}$. In fact, as the MILP does not distinguish between $\rho$-$MMS$ allocations and $\rho$-$PMM$ allocations (with $U$ being the agent for which we consider the proportional share), we get the same value of $\rho$ also for $\rho$-$PMM$ allocations. (The MMS partition of agent $U$ has no effect on the atomic bundles. Consequently, we may without loss of generality assume that her MMS equals her PS, as each atomic bundle may be composed of three items of identical value for $U$.) For $\rho$-PMM allocations, our results are tight. The MILP generates explicit instances for which there is no $\rho$-PMM allocation with $\rho > \frac{11}{12}$. 

\begin{theorem}
\label{thm:PMM}
For three agents with additive valuations over goods, coarse atomic exhaustive search generates a $\rho$-$PMM$ allocation with $\rho \ge \frac{11}{12}$ (and hence also a $\rho$-$MMS$ allocation with $\rho \ge \frac{11}{12}$). Moreover, given the MMS partitions of agents $R$ and $C$, the allocation algorithm runs in polynomial time.
\end{theorem}

\begin{proposition}
\label{pro:PMM}
There are instances with three agents (named $R$, $C$ and $U$)  with additive valuations over goods for which in every allocation, either $R$ gets at most a $\frac{11}{12}$ fraction of her MMS, or $C$ gets at most a $\frac{11}{12}$ fraction of her MMS, or $U$ gets at most a $\frac{11}{12}$ fraction of her PS. 
\end{proposition}

\subsection{Ambitious coarse atomic exhaustive search}
\label{sec:9Gvariation}

We extended the MILP approach to analyse a more ambitious version of coarse atomic exhaustive search, in which the algorithm searches for an atomic allocation in which $U$ gets at least her full proportional share, whereas $R$ and $C$ get at least $\rho$ times their MMS, for $\rho$ as large as possible. The natural way of generating such an extension involves introducing constraints with strict inequalities in the associated MILP, whereas MILPs require constraints with weak inequalities. We introduce a slackness variable that allows us to replace strict inequalities by weak inequalities, initialize its value to a fixed constant ($\frac{1}{522}$), and provide analysis that shows that the new MILP that has only weak inequalities is equivalent to the original MILP that had also strong inequalities. Using this MILP, we determine that ambitious coarse atomic exhaustive search also guarantees a $\rho$-$P^+MM$ allocation with $\rho = \frac{11}{12}$. 

\begin{theorem}
\label{thm:P+MM}
For three agents with additive valuations over goods, ambitious coarse atomic exhaustive search generates a $\rho$-$P^+MM$ allocation with $\rho \ge \frac{11}{12}$. Moreover, given the MMS partitions of agents $R$ and $C$, the allocation algorithm runs in polynomial time.
\end{theorem}

Observe that Theorem~\ref{thm:P+MM} implies Theorem~\ref{thm:PMM}, and Proposition~\ref{pro:PMM} implies that both these theorems are tight.
We do not know whether it is ``coincidental" that the best values of $\rho$ for $\rho$-$PMM$ and $\rho$-$P^+MM$ allocations are the same, or whether there is some underlying principle from which such an equality can be deduced.

\subsection{Chores}
\label{sec:9C}

Similar to the case of goods, one can consider the corresponding coarse atomic exhaustive search algorithm for chores, and analyse its performance using an associated MILP. 
The results obtained for chores are analogous to those obtained for goods, but with a different value of $\rho$. 

\begin{theorem}
\label{thm:P+MMC}
For three agents with additive valuations over chores, ambitious coarse atomic exhaustive search generates a $\rho$-$P^+mm$ allocation with $\rho \le \frac{19}{18}$ (and hence also a $\rho$-$mmS$ allocation with $\rho \le \frac{19}{18}$). Moreover, given the mmS partitions of agents $R$ and $C$, the allocation algorithm runs in polynomial time.
\end{theorem}

\begin{proposition}
\label{pro:PMMC}
There are instances with three agents (named $R$, $C$ and $U$)  with additive valuations over chores for which in every allocation, either $R$ gets at least $\frac{19}{18}$ times the cost of her mmS, or $C$ gets at least $\frac{19}{18}$ times the cost of her mmS, or $U$ gets at least $\frac{19}{18}$ times the cost of her PS. 
\end{proposition}

Obtaining the example proving Proposition~\ref{pro:PMMC} was more difficult than obtaining the example proving Proposition~\ref{pro:PMM}. The MILP that was used in order to prove Theorem~\ref{thm:PMM} (this MILP is omitted from this paper as we present the MILP for the stronger Theorem~\ref{thm:P+MM}) generated the example for Proposition~\ref{pro:PMM}. However, the corresponding MILP for chores did not generate an example that proves Proposition~\ref{pro:PMMC}. (So as to be able to solve the MILPs in practice, the MILPs that we use do not contain all constraints that are implied by the corresponding allocation algorithm, but rather a subset of them that suffices in order to get the correct value for the objective function. The MILP generates negative examples that do not violate any of the constraints that it uses, but these negative examples might violate constraints that were discarded.) Consequently, generating the example proving Proposition~\ref{pro:PMMC} required extra work on our behalf.

\subsection{Coarse atomic partial search}
\label{sec:9P}

In passing, we revisit the algorithm that gave rise to the previous best value of $\rho$ known for $\rho$-MMS allocations for goods when $n=3$.  This algorithm was introduced in~\cite{AMNS17} where a bound of $\rho\ge \frac{7}{8}$ was proved, and its analysis was improved in~\cite{GM19}, showing that  $\rho\ge \frac{8}{9}$, and providing an example showing that for this particular algorithm $\rho$ is no better than $\frac{11}{12}$. We refer to that algorithm as {\em coarse atomic partial search}. The algorithm proceeds as follows, with agents named $R$, $C$ and $U$, as in coarse atomic exhaustive search.

\begin{enumerate}
    \item If there are two different bundles in the MMS partition of $C$ (call this partition $(C_1, C_2, C_3)$), one giving $R$ at least her MMS and the other giving $U$ at least her PS, then give each of $R$ and $U$ the corresponding bundle, and allocate the remaining bundle to $R$. This gives a PMM allocation. Hence we proceed under the assumption that there are no such two bundles. This implies that exactly one ``valuable" bundle among $(C_1, C_2, C_3)$  simultaneously gives $R$ at least her MMS and gives $U$ at least her PS, and each of the remaining two bundles fail to give $R$ her MMS and fail to $U$ her PS. By renaming, we may assume that $C_1$ is the valuable bundle. 
    \item As in coarse atomic exhaustive search, create nine atomic bundles, by considering the intersections of the MMS bundles of $R$ and $C$.
    \item Consider all atomic allocations $(A_1, A_2, A_3)$ in which $A_1$ is either $C_2$ or $C_3$. Among them, choose the allocation that maximizes $\min[v^R(A_2), v^R(A_3)]$. In this allocation, give $A_1$ to $C$ (thus $C$ gets at least her MMS), of the remaining two bundles, give $U$ the bundle of higher value to $U$ (thus $U$ gets at least her $PS$, as $v^U(A_1) \le \frac{1}{3}v^U(\items)$), and give $R$ the bundle that remains ($R$ gets at least a $\rho$ approximation to her MMS). 
\end{enumerate}

An advantage of coarse atomic partial search over coarse atomic exhaustive search is that the MILP for analysing it has much fewer variables. It suffices to analyse only steps~3 of the algorithm, and for that, only the valuation function of $R$ matters (this gives nine real variables, one for each atomic bundle), and no variables need to be introduced in order to account for the valuation functions of $C$ and $U$. Previous work~\cite{AMNS17,GM19} attempted to analyse this algorithm by hand, and failed to achieve tight results. We analyse it by setting up the corresponding MILP and running it, proving that $\rho \ge \frac{9}{10}$, and generating an allocation instance for which $\rho = \frac{9}{10}$.

Observe that coarse atomic partial search produces an allocation that we may refer to as a $\rho$-$P^+M^+M$ allocation: agent $U$ gets at least her PS, agent $C$ gets at least her MMS, and agent $R$ gets at least a $\rho$ fraction of her MMS. However, the value of $\rho$ that it gives ($\frac{9}{10}$) is not as good as the one  for $\rho$-$P^+MM$ allocations (which is $\frac{11}{12}$, by coarse atomic exhaustive search).

\begin{theorem}
\label{thm:P+M+M}
For three agents with additive valuations over goods, coarse atomic partial search produces a $\rho$-$P^+M^+M$ allocation with $\rho \ge \frac{9}{10}$. Moreover, given the MMS partitions of agents $R$ and $C$, the allocation algorithm runs in polynomial time.
\end{theorem}

\begin{proposition}
\label{pro:P+M+M}
There are instances with three agents (named $R$, $C$ and $U$)  with additive valuations over goods for which in the allocation produced by coarse atomic partial search, $R$ gets at most a $\frac{9}{10}$ fraction of her MMS. 
\end{proposition}

Our results for coarse atomic partial search demonstrate that even for allocation algorithms that are much simpler to analyse than the main algorithm considered in our work (which is coarse atomic exhaustive search), analysis that is not computer assisted (as attempted in~\cite{AMNS17,GM19}) failed to determine the correct value of $\rho$. This suggests that computer assisted analysis, and specifically, the use of MILPs, might be necessary (though unfortunately, maybe not sufficient) in order to obtain tight or nearly tight bounds of the best values of $\rho$ for which $\rho$-$MMS$ allocations exist. 

We leave the question of what is the best value of $\rho$ in  $\rho$-$P^+M^+M$ allocations open. (Proposition~\ref{pro:P+M+M} applies to coarse atomic partial search, but not to $\rho$-$P^+M^+M$ allocations in general.)

\section{Analysis of the allocation algorithms}
\label{sec:MILP}

In this section we prove theorems and propositions that were presented without proof in Section~\ref{sec:algorithms}.

\subsection{Atomic exhaustive search}
\label{sec:atomic}

We prove here Theorem~\ref{thm:AES}, that for every $n$, the approximation ratio $\rho_n$ guaranteed by atomic exhaustive search can be computed by solving a mixed integer/linear program (MILP) of finite size. (The integer variables take only 0/1 values, and hence the corresponding MILP is in fact a mixed Boolean/linear program.)

\begin{proof}
As $\rho_2=1$ (the cut and choose procedure gives each of the two agents her maximin share), we may assume that $n\ge 3$.

We wish to construct an allocation instance with $n$ agents with additive valuations over $n^n$ atomic bundles (where these atomic bundles correspond to intersections of bundles of MMS partitions of the agents), in which in every allocation there is an agent that gets a bundle of value no more than $\rho$ times her MMS. Among all such instances, we wish to find the one with the smallest value of $\rho$, and we denote this value by $\rho_n$. To simplify terminology and without loss of generality, we may assume that each atomic bundle is composed of a single item.

Without loss of generality we may further assume that in such an instance, the MMS of every agent is strictly positive. Consequently, all valuation functions can be scaled to give the same $MMS$ value. We denote this value as $z+1$.
We want some agent in every allocation to get a value of at most $z$, which would mean that $\rho = \rho(z)=\frac{z}{z+1}.$ As $\rho(z)$ monotonically increases in $z,$ we shall minimize $z$.

Our program has the following variables:

\begin{enumerate}
    \item $x_{ij}$: the value of item $j$ to agent $i$, for every $i$ and $j$.
    \item $z$: the value such that in every allocation at least one agent does not get value larger than $z$. The MMS of every agent is $z+1$.
    \item $y_{iB}:$ a binary variable for every agent $i$ and bundle $B$, with $y_{iB} = 1$ indicating that $v^i(B) \le z$.
\end{enumerate}

The objective function of the program is to minimize $z$, subject to the following constraints:

\begin{enumerate}
    
    \item $x_{ij}\ge 0$ for every $i$ and $j$.
    
    \item $z\ge  0$.
    
    \item For every agent $i$ and bundle $B$ in $i$'s MMS partition, $\sum_{j \in B} x_{ij} = z+1$. 
    
    \item For every agent $i$ and bundle $B$, we have the constraint $v^i(B)\le  z + \gamma(1-y_{iB})$.  If $y_{iB}=1$, then $v^i(B)\le  z$. If $y_{iB}=0$, then $v^i(B)$ is practically not constrained. (A value of $\gamma$ that suffices for $v^i(B)$ not to be constrained is determined as follows. In~\cite{FST21} it is shown that $\rho_n\le  1-\frac{1}{n^4}$ for every $n \ge  3$, which implies that $z\le  n^4-1.$ This implies that $\gamma=n^5$ suffices.)
    
    \item For every allocation $A=(A_1,\ldots,A_n)$ we have the constraint $\sum\limits_i y_{iA_i}\ge 1$, which implies that at least one agent gets value at most $z$.

\end{enumerate}

The above algorithm defines an MILP. As argued above, the MILP is feasible with a value of $z \le n^4$. As we have the constraint $z \ge 0$, it is also bounded (observe that the value of every $x_{ij}$ variable is bounded between~$0$ and~$z+1$). Being an MILP with integer coefficients, this implies that it has an optimal solution that is rational. (For each one of the finitely many ways of assigning 0/1 values for the binary variables $y_{iB}$, we get an LP with integer coefficients. We only minimize over those LPs that are feasible, and among those, we only care about those LPs of value at most $n^4$. Each such LP has an optimal solution that is rational.) Consequently, $\rho_n = \frac{z}{z+1}$ exists and is rational as well.
\end{proof}

\subsection{Terminology and conventions for coarse atomic bundles}
\label{sec:terminology}

In the rest of this manuscript, we only consider allocation instances with $n=3$ agents. 
The agents are named as $R$, $C$ and $U$ (we may arbitrarily choose which agent gets each name). We consider the MMS partitions of $R$ and of $C$, but not of $U$. The atomic bundles are the nine intersections, one bundle from each partition.  Rearranging them in a three by three matrix, we number them as follows:

$\left(
  \begin{array}{ccc}
    e_1 & e_2 & e_3 \\
    e_4 & e_5 & e_6 \\
    e_7 & e_8 & e_9 \\
  \end{array}
\right)$

For $R$ the MMS partition is the rows and for $C$ it is the columns. Observe that $U$'s MMS partition might require breaking some of the atomic bundles, and hence it is not represented in the resulting MILP.

The rows and columns of the matrix are denoted by $r_1, r_2, r_3$ (from top to bottom) and $c_1, c_2, c_3$ (from left to right). The MMS of $R$ and $C$ and the PS of $U$ are denoted $z+1$. Without loss of generality we may assume that:

\begin{itemize}

    \item $v^R(r_j) = z+1$ for every $j \in \{1,2,3\}$. 
    \item $v^C(c_j) = z+1$ for every $j \in \{1,2,3\}$.
    \item $v^U(\items) = 3z + 3$ (by definition).
\end{itemize}

(If any constraint in the first two sets of constraints above does not hold, then we can reduce values of items without reducing the MMS or PS of the agents, and the value of $\rho_A$ for any allocation $A$ does not improve.)

For justification of the last two sets of constraints, see the first step of the allocation algorithm coarse atomic partial search, explained in Section~\ref{sec:9P}. The constraints come in two variations, one for $\rho$-$PMM$ allocations, and one for $\rho$-$P^+MM$ allocations.

\begin{itemize}
    
    \item $v^C(r_2) \le  z$,\; $v^C(r_3) \le  z$,\; $v^R(c_2) \le  z$,\; $v^R(c_3) \le  z$.  
    \item For $\rho$-$PMM$ allocations: $v^U(r_2) \le  z$,\; $v^U(r_3) \le  z$,\;  $v^U(c_2) \le  z$,\; $v^U(c_3) \le  z$.
    \item For $\rho$-$P^+MM$ allocations: $v^U(r_2) < z+1$,\; $v^U(r_3) < z+1$,\;  $v^U(c_2) < z+1$,\; $v^U(c_3) < z+1$.
    
\end{itemize}

\subsection{Negative example for PMM allocations}

Here we prove Proposition~\ref{pro:PMM}, presenting an instance for which no $\rho$-$PMM$ allocation has $\rho > \frac{11}{12}$.  

\begin{proof}
Consider an instance with nine items (corresponding to the coarse atomic bundles) and the following three valuation functions (depicted according to the conventions of Section~\ref{sec:terminology}):

$v^R =\left(
  \begin{array}{ccc}
    0   & 6 & 6 \\
    10  & 1 & 1 \\
    4   & 4 & 4 \\
  \end{array}
\right),$

$v^C =\left(
  \begin{array}{ccc}
    0               &   7+\frac{1}{3}   &   6+\frac{2}{3} \\
    8+\frac{1}{3}   &   1               &   1+\frac{2}{3} \\
    3+\frac{2}{3}   &   3+\frac{2}{3}   &   3+\frac{2}{3} \\
  \end{array}
\right),$

$v^U =\left(
  \begin{array}{ccc}
    0               &   7+\frac{1}{3}   &   7+\frac{1}{3}   \\
    10+\frac{1}{3}  &   0               &   0               \\
    3+\frac{2}{3}   &   3+\frac{2}{3}   &   3+\frac{2}{3}   \\
  \end{array}
\right).$

Observe that in the above example, the MMS of each of the agents $R$ and $C$ is~12, and so is the PS of $U$.

In every allocation of the nine items to the three agents, at least one agent gets a bundle of value at most~$11$. This can be seen by a case analysis, that breaks into cases depending on which agent gets item $e_4$ (the item of largest value in this instance).
\begin{enumerate}
    \item Suppose that agent $R$ takes $e_4$. For $U$ to get the sum larger than $11$, she must take at least two items valued $7+\frac{1}{3}$ or one item valued $7+\frac{1}{3}$ and at least two items valued $3+\frac{2}{3}.$ In the first case, $C$ must take the third row and at least one of the remaining non-zero items, as otherwise it would be insufficient for her to get the value larger than $11$. However, that leaves $R$ with the sum at most $11$. In the second case, $C$ is left with the items $e_2, e_5, e_6,$ and, say, $e_9$ (as $e_7=e_8=e_9$ for any agent). $C$ must take $e_2$ and $e_9$ and one of the rest to get the sum larger than $11$. However, that leaves $R$ with $11$ at most.
    \item Suppose now that agent $C$ takes $e_4$. Again, for $U$ to get the sum larger than $11$, she must take at least two items from the first row, that is, $e_2$ and $e_3$, or at least two items from the third row and one item from the first row, - for instance, $e_7, e_8$, and $e_3$. In the first case, $R$ must take at least the third row. However, this leaves $C$ with at most $e_5$ and $e_6$, which give him $11$ when added to $e_4.$ In the second case, $R$ must take all the rest items in order to get $12$ (since $R$ is integer valued, she aims to get at least $12$). That leaves $C$ with $e_4$ only, which is too small.
    \item Finally, we suppose that agent $U$ takes $e_4$. It suffices for $U$ to take $e_7$ (and this is the lowest value item that would suffice). Now for $R$ to get at least $12$ it is suffices to take at least first row, or to take elements $e_3, e_5, e_6,$ and, say, $e_8$. Those scenarios allow for $C$ to get at most $10$ or $11$, respectively. 
\end{enumerate}

Therefore, in the instance above there is no $\rho$-PMM allocation with $\rho > \frac{11}{12}.$
\end{proof}

Observe that for the instance presented in the proof of Proposition~\ref{pro:PMM}, item $e_1$ has value~0 for all agents. Hence effectively, the instance has only eight items. As proved in~\cite{FST21}, for every instance with eight items (and additive valuations), there is an MMS allocation. Indeed, this holds for the instance of Proposition~\ref{pro:PMM}. The MMS of $U$ is~11, and the allocation $A = \{A_R = (e_1,e_4,e_6,e_7), A_C = (e_2, e_5, e_8), A_U = (e_3, e_9)\}$ gives the agents values $15$, $12$ and $11$, respectively. 

\subsection{Ambitious coarse atomic exhaustive search}

In this section we prove Theorem~\ref{thm:P+MM}, that ambitious coarse atomic exhaustive search produces a $\rho$-$P^+MM$ allocation with $\rho \ge \frac{11}{12}$. For this purpose, we use an MILP that we call MILP9G (as it concerns goods and nine atomic bundles).

Naively, MILP9G would requires $3\cdot 9 + 1=28$ real variables, around $3 \cdot 2^9 \simeq 1500$ binary variables, and around $3^9 \simeq 20000$ linear constraints. However, as we shall see it, a much smaller subset of constraints will suffice in order to find the optimal solution for MILP9G, and consequently also most of the binary variables will not be needed.

Initially, we present a formulation of MILP9G in which some of the constraints involve strict inequalities. We refer to this formulation as MILP9g (even though MILPs are not supposed to contain strict inequalities). Later we shall show how to replace the strict inequalities of MILP9g by weak inequalities, thus obtaining MILP9G. 

MILP9g has the following variables.

\begin{enumerate}

\item $e^R_j$, $e^C_j$ and $e^U_j$, for every item $j$: the value of item $j$ to the respective agent.

\item $z$: the value such that in every allocation, at least one agent does not get value above $z$. 

\item For some of the bundles $B$, as listed below, we have binary variables $R_{B}$, $C_{B}$ and $U_{B}$. Setting either $R_B$ or $C_B$ to~1 implies that bundle $B$ has value at most $z$ for the respective agent, and if set to~0 the value of bundle $B$ to the agent is not constrained. Likewise, setting $U_B$ to~1 implies that bundle $B$ has value less than $z+1$ for agent $U$ (namely, $\sum_{j\in B} e_j^U < z+1$). The bundles $B$ are only of the following types:

\begin{enumerate}

\item Agent $R$ gets a single row $r_2$ or $r_3$. Agent $C$ gets either one or two items from row $r_1$ and maybe some other items, and all remaining items are allocated to agent $U$.

\item Agent $C$ gets a single column $c_2$ or $c_3$. Agent $R$ gets either one or two items from column $c_1$ and maybe some other items, and all remaining items are allocated to the agent $U$. 

\end{enumerate}

\end{enumerate}

The objective function is to minimize $z$, subject to the following constraints: 

\begin{enumerate}

\item $e^R_j, e^C_j, e^U_j \ge  0$ (for every item $e_j$ for each agent). 

\item Recalling the naming of rows are columns, we have the linear constraints stating that $v^R(r_j) = z+1$ for every row $r_j$, and that $v^C(c_j) = z+1$ for every column $c_j$. For $U$, we have the constraint that the value of the grand bundle is $3z+3$.

\item Recalling Section~\ref{sec:terminology}, we have the constraints $v^C(r_2) \le  z$, $v^C(r_3) \le  z$, $v^R(c_2) \le  z$, $v^R(c_3) \le  z$, and the constraints $v^U(r_2) < z+1$, $v^U(r_3) < z+1$, $v^U(c_2) < z+1$, $v^U(c_3) < z+1$. 

\item For each agent $Y \in \{R,C\}$ and bundle $B$ 
we have the constraint $v^Y(B) \le  z  + \gamma(1 - Y_B)$, where $\gamma$ is a sufficiently large constant ($\gamma = 120$ suffices, as negative examples of~\cite{FST21} imply that MILP9g is feasible with $z=39$). For agent $U$, the corresponding constraint is $v^U(B) < z - 1 + \gamma(1 - R_{B}).$

\item We consider only allocations $A = (A_R, A_C, A_U)$ of the following four types:

\begin{enumerate}

\item $A_R = r_2$.

\item $A_R = r_3$. 

\item $A_C = c_2$.

\item $A_C = c_3$. 

\end{enumerate}

For each such allocation, we have a linear constraint that says that the sum of the two binary variables that represent the other two bundles is at least~1. This gives $4 \cdot 6 \cdot 8 = 192$ constraints. (E.g., if $A_R = r_2$, then $A_C$ must contain either one or two items from $r_1$, giving~6 possibilities, and for any such choice, and number of items from $r_3$, giving eight possibilities.) For readability, these constraints are ordered and grouped together in a systematic way so that it is easy to see that the program contains the correct constraints.

\end{enumerate}

MILP9g is feasible. In particular, the negative example given in \cite{FST21} translates into a feasible solution for MILP9g with an objective value of $z=39$. 

In order to modify MILP9g into MILP9G, we replace the strict inequalities of the form $< z+1$ by weak inequalities of the form $\le z+1-\epsilon$, for some small $\epsilon<1$. 
The following proposition determines a sufficiently small value for $\epsilon$.

\begin{proposition}
For $\epsilon=\frac{1}{522},\ z$ is feasible for MILP9g if and only if $z$ is feasible for MILP9G.
\end{proposition}

\begin{proof}
Every feasible solution for MILP9G is a feasible solution for MILP9g, because the constraints of MILP9G are stricter than those of MILP9g. We now prove the other direction.

Let $z$ be feasible for  MILP9g, and consider an arbitrary feasible solution $X=\{e^R_i, e^C_j, e^U_k, y_{\ell}, z\}$ (where the $y_{\ell}$ are the values of te binary variables) of MILP9g with value $z$. We modify $X$ into a solution $X'$ with the same value of $z$, where $X'$ is feasible for  MILP9G. In $X'$ we keep the values of most variables as they are in $X$, except for the values of variables $e^U_1,\ldots,e^U_9$.  These values will be changed so as to make the values of all strict inequalities (of the form $< z+1$) satisfied with a margin as large as possible (this margin corresponds to $\epsilon$). We now explain how to compute a feasible $\epsilon$.
For this purpose, we construct an auxiliary linear program.

The auxiliary linear program (ALP) contains the real variables $e^U_1,\ldots,e^U_9$ of MILP9b$^1$, which for simplicity we rename as $e_1,\ldots, e_9$. ALP also contains those constraints from MILP9b$^1$ that involve the variables $e^U_1,\ldots,e^U_9$, (but not the constraints that involve $e_i^R$ and $e_i^C$). One set of constraints is thus $e_i\ge 0$ for every $i$. An additional change, done for simplicity of notation, is to scale the value $z$ (which is a value, not a variable, since it is taken from the fixed $X$), by a multiplicative factor of $\frac{1}{z+1}$ (and, consequently, values of other real variables are scaled by the same factor). Hence, the constraint $\sum\limits_i e_i = 3z+3$ in MILP9g is replaced by $\sum\limits_i e_i=3$ in ALP. Finally, we introduce a new variable $w$ that represents an upper bound (strictly smaller than $1$) on the values of those bundles $B$ for which the solution $X$ dictated that the value for agent $U$ of bundle $B$ must be strictly smaller than $z+1$ (that is, in $X$ the binary variable $U_B$ associated with bundle $B$ has value~1). Consequently, we have the four constraints $\sum\limits_{i\in S} e_i - w \le  0,$ where $S$ is either $r_2, r_3, c_2,$ or $c_3$; and in addition, a list of constraints of the form $\sum\limits_i e_i - w \le  0$, for each bundle $B$ as above. The objective is to minimize $w$. 

Let $r$ denote the number of constraints that we have. We can express the inequalities of ALP in a matrix form $A x\le  b$, where $A$ is a matrix of size $r\times10$, $x$ is a vector of ten variables, and $b$ is a vector of dimension $r$. 
$$\left(
\begin{array}{cccc}
    1 & \ldots & 1 & 0  \\
    \ldots & \ldots & \ldots & -1 \\
    \vdots & \ddots & \ddots &\vdots \\
    \ldots & \ldots & \ldots & -1
\end{array}
\right)\left(
\begin{array}{c}
e_1\\
\vdots\\
e_9\\
w
\end{array}
\right)\le  \left(
\begin{array}{c}
3\\
0\\
\vdots\\
0
\end{array}
\right).$$

Since ALP is both feasible and bounded, its optimal solution is obtained at a basic feasible solution (BFS). 
The BFS is a solution to a linear system of equations $A' x' = b'$. For some $k \le 10$, $x'$ is a vector containing $k$ of the variables of $x$, $A'$ is a $k$ by $k$ invertible matrix whose rows are composed of $k$ of the constraints of $A$ (and columns of variables not in $x'$ are deleted), and $b'$ contains only the $k$ entries of $b$ that correspond to the constraints in $A'$. In this BFS, the values of those variables of $x$ not contained in $x'$ is~0. Necessarily, $x'$ contains the variable $w$ (as its value in the optimal solution for ALP is positive). Likewise, $A'$ contains the first row of $A$ (and $b'$ contains the first entry of $b$), as otherwise $x' = 0$ solves $A' x' = b'$.
Hence $A' x' = b'$ can be depicted as follows:

$$\left(
\begin{array}{cccc}
    1 & \ldots & 1 & 0  \\
    \ldots & \ldots & \ldots & -1 \\
    \vdots & \ddots & \ddots &\vdots \\
    \ldots & \ldots & \ldots & -1
\end{array}
\right)\left(
\begin{array}{c}
e_{i_1}\\
\vdots\\
e_{i_{k-1}}\\
w
\end{array}
\right)=\left(
\begin{array}{c}
3\\
0\\
\vdots\\
0
\end{array}
\right).$$

We can solve the {system of equations} to find $w$ with the use of the Cramer's rule. That is,

$$w=\frac{\det(A'_w)}{\det (A')},$$
where $A'_w$ is the matrix $A'$ with its last column being substituted by $b'$. The matrix $A'$ is a $(-1,0,1)$-matrix. We observe that $-1$ appears only in its last column, which contains no $+1$ entry. Denoting by $|A'|$ the matrix $A'$ with the sign of the last column changed, we get that $\det A' =-\det |A'|$. Now, since both numerator and denominator are integer numbers, and $w<1$, it holds that %
$$w\le 1-\frac{1}{|\det |A'||}.$$

Since $|A'|$ is a $(0,1)$-matrix of order $k$, we can apply the following upper bound (see \cite{FS65}, problem 523):
$$\det |A'|\le \frac{(k+1)^\frac{(k+1)}{2}}{2^k},$$
This upper bound is similar in spirit to Hadamard's upper bound for the determinant of $(-1, 1)$-matrices.

For $k \le 10$, the upper bound function is maximized at $k=10$. Applying the inequality to $w$, in the worst case we get:

$$w\le  1-\frac{1}{|\det |A| |}\le  1 - \frac{2^{10}}{11^{11/2}}  \le 1-\frac{1}{522}.$$

Hence ALP has a feasible solution with a value of $w$ as above. Scaling the solution of ALP by a multiplicative value of $z+1$, we modify the solution $X$ for MILP9g by replacing the values of the variables $e^U_1,\ldots,e^U_9$ by the scaled values of $e_1, \ldots, e_9$, thus obtaining $X'$. The new solution $X'$ is indeed feasible for MILP9G, because all the strict inequalities of the form $< z+1$ that were satisfied by $X$ in MILP9g, are satisfied by $X'$ as weak inequalities of the form $\le (z+1)w $ (and this last value is at most $z + 1 - \frac{1}{522}$, because $z \ge 0$ and $w \le 1-\frac{1}{522}$).
\end{proof}

MILP9G was solved by an MILP solver (it took around 70 seconds for https://online-optimizer.appspot.com/ to solve MILP9G), with the result being $\rho=\frac{11}{12}$. The code for MILP9G is presented in Appendix~\ref{app:milp9b} and a C++ program that generates MILP9G is presented in Appendix~\ref{app:9bprogramGoods}. 

\subsection{A negative example for chores}

In this section we prove Proposition~\ref{pro:PMMC}, presenting an instance in which no $\rho$-$Pmm$ allocation (for chores) has $\rho < \frac{19}{18}$.
\begin{proof}
Consider an instance with nine items (corresponding to the coarse atomic bundles) and the following three valuation functions (depicted according to the conventions of Section~\ref{sec:terminology}):

$v^R(M)=\left(
  \begin{array}{ccc}
    0   & 9 & 9 \\
    12  & 3 & 3 \\
    4   & 7 & 7 \\
  \end{array}
\right),$

$v^C(M)=\left(
  \begin{array}{ccc}
    0               &   8.5   &   7.5 \\
    12.75   &  3.25   &   3           \\
    5.25   &   6.25  &   7.5 \\
  \end{array}
\right),$

$v^U(M)=\left(
  \begin{array}{ccc}
    0               &   7.6   &   7.6   \\
    11.4 &   4.6   &   3.8               \\
    3.8  &   7.6   &   7.6  \\
  \end{array}
\right).$

In the above example, the MMS of each of the agents $R$ and $C$ is~18, and so is the PS of $U$.

In every allocation of the nine items to the three agents, at least one agent gets bundle of value at least~$19$.
This can be seen by a case analysis, breaking in cases according to the agent that receives $e_4$ (the chore of highest cost). 

\begin{enumerate}
    \item Suppose that agent $R$ takes $e_4$. For her to get the sum below $19$, she can take either $e_7$ or both $e_5$ and $e_6$. Suppose that $R$ takes $e_7$. $U$ can take either two items valued $7.6$, or $e_5$, $e_6$, and any other non-zero element. In either case, $C$ is left with the sum no less than $20$ or $21.25$, respectively. Suppose now that $R$ takes $e_5$ and $e_6$ instead. $U$ can take either two items valued $7.6$, or $e_7$ and and any other non-zero item. In either case, $C$ is left with the sum no less than $19$ or $21.25$, respectively.
    \item Suppose now that agent $C$ takes $e4$. For her to get the sum below $19$, she can take either $e_5$ or $e_7$ (there is no point in taking $e_6$ as all agents value it less than $e_5$). Suppose that $C$ takes $e_5$. $U$ can take either two items valued $7.6$, or $e_6, e_7$, and any other non-zero element. In either case, $R$ is left with the sum no less than $21$ or $23$, respectively. 
    Suppose that $C$ takes $e_7$. Then $U$ can take either two items valued $7.6$, or $e_5, e_6$ and any other non-zero element. In either case, $R$ is left with the sum no less than $20$ or $23$, respectively.
    \item Finally, suppose that agent $U$ takes $e4$. For her to get the sum below $19$, she can take $e_7$ (the items $e_5$ and $e_6$ have lower value to the other agents, so there is no point in taking them instead).
    The sum of any four non-zero items of agent $R$ is at least $19$, and likewise for $C$. This  implies that both $R$ and $C$ must have exactly three non-zero items. As $e_2$ is the largest item for the two, we consider two cases:
    \begin{enumerate}
        \item Suppose that $R$ takes $e_2$. She cannot take any items but $e_5$ and $e_6$. That leaves $C$ with the sum $21.25$.
        \item Suppose that $C$ takes $e_2$. She can additionally take either $e_5$ and $e_8$, $e_5$ and $e_6$, or $e_8$ and $e_6$. In either case, $R$ is left with either $19, 25$, or $19$, respectively.
    \end{enumerate}
\end{enumerate}

Therefore, in this example there is no $\rho$-Pmm allocation for chores with $\rho < \frac{19}{18}.$
\end{proof}

\subsection{Analysis of allocation algorithm for chores}

As noted in Proposition~\ref{pro: minimax equivalence}, the $\rho$-maximin problem for chores can be reformulated as a $\rho$-minimax problem for goods. This change is performed to conveniently avoid the work with negative numbers. It is implied by the definition that in case of minimax $\rho\ge 1$.

In this section we prove Theorem~\ref{thm:P+MMC}, that ambitious coarse atomic exhaustive search produces a $\rho$-$P^+mm$ allocation with $\rho \le \frac{19}{18}$. Our proof is based on running an MILP that we refer to as MILP9C.

As in the case of MILP9G, we fix the agent who is to get her full $PS$ as $U$. We also scale every agent's valuation function to get the same $mmS$ value, which we denote as $z-1.$ We want some agent to get a value at least $z$, which means that $\rho=\frac{z}{z-1}.$ As $\rho(z)$ monotonically decreases in $z$, we shall minimize $z$.

MILP9C has the following variables:

\begin{enumerate}

\item $e^R_j$, $e^C_j$ and $e^U_j$, for every item $j$: the value of item $j$ to the respective agent.

\item $z$: the value such that in very allocation, at least one agent does not get value less than $z$. 

\item For some of the bundles $B$, as listed below, we have binary variables $R_{B}$, $C_{B}$ and $U_{B}$. Setting either $R_B$ or $C_B$ to~1 implies that bundle $B$ has value at least $z$ for the respective agent, and if set to~0 the value of bundle $B$ to the agent is not constrained. Likewise, setting $U_B$ to~1 implies that bundle $B$ has value more than $z-1$ for agent $U$ (namely, $\sum_{j\in B} e_j^U < z-1$). The bundles $B$ are only of the following types:

\begin{enumerate}

\item Agent $R$ gets a single row $r_2$ or $r_3$. Agent $C$ gets either one or two items from row $r_2$ or $r_3$, whichever is left, and maybe some other items, and all remaining items are allocated to agent $U$.

\item Agent $C$ gets a single column $c_2$ or $c_3$. Agent $R$ gets either one or two items from column $c_2$ or $c_3$, whichever is left, and maybe some other items, and all remaining items are allocated to agent $U$. 

\end{enumerate}

\end{enumerate}

The objective function is to minimize $z$, subject to the following constraints: 

\begin{enumerate}

\item $e^R_j, e^C_j, e^U_j \ge  0$ (for every item $e_j$ for each agent). 

\item Recalling the naming of rows are columns, we have the linear constraints stating that $v^R(r_j) = z-1$ for every row $r_j$, and $v^C(c_j) = z-1$ for every column $c_j$. For $U$, we have the constraint that the value of the grand bundle is $3z-3$.

\item Recalling Section~\ref{sec:terminology}, we have the constraints $v^C(r_2) \ge  z$, $v^C(r_3) \ge  z$,  $v^R(c_2) \ge  z$, $v^R(c_3) \ge  z$, and the constraints $v^U(r_2) > z-1$, $v^U(r_3) > z-1$, $v^U(c_2) > z-1$, $v^U(c_3) > z-1$. 

\item For each agent $Y \in \{R,C\}$ and bundle $B$ we have the constraint $v^Y(B) \ge  z  - \gamma(1 - Y_B)$, where $\gamma$ is a sufficiently large constant ($\gamma = 44$ suffices, as negative examples of~\cite{FST21} imply that MILP9g is feasible with $z=44$). For agent $U$, the corresponding constraint is $v^U(B) > z - 1 - \gamma(1 - R_{B}).$ 

\item We consider only allocations $A = (A_R, A_C, A_U)$ of the following four types:

\begin{enumerate}

\item $A_R = r_2$.

\item $A_R = r_3$. 

\item $A_C = c_2$.

\item $A_C = c_3$. 

\end{enumerate}

For each such allocation, we have a linear constraint that says that the sum of the two binary variables that represent the other two bundles is at least~1. This gives $4 \cdot 6 \cdot 8 = 192$ constraints. For readability, these constraints are ordered and grouped together in a systematic way so that it is easy to see that the program contains the correct constraints.

\end{enumerate}

{
MILP9C has strict inequalities, whereas to solve MILPs we would like it to have only weak inequalities. MILP9C is feasible (the negative example provided in~\cite{FST21} implies that it has a feasible solution with $z= 44$).  Similarly to MILP9g, we can modify MILP9C by replacing the strict inequalities with weak inequalities (that is, $>z-1$ is replaced by $\ge  z-1+\frac{1}{522}$). Correctness of the modification is proved in the same way as it was proved for the case of goods, and thus it is not repeated.

MILP9C was solved by an MILP solver (in took around 210 seconds to solve MILP9C using https://online-optimizer.appspot.com/), with the result being $\rho=\frac{19}{18}$. The code for MILP9C, as well as a C++ program that generates MILP9C, can be obtained upon request from the authors.
}

\subsection{Analysis of coarse atomic partial search}

We prove Proposition~\ref{pro:P+M+M}, presenting an example in which coarse atomic partial search achieves a value of $\rho$ no better than $\frac{9}{10}$.

\begin{proof}
The values of the coarse atomic bundles for agent $R$ are depicted below:

$\left(
  \begin{array}{ccc}
    6    & 3 & 1 \\
    6    & 3 & 1 \\
    0   & 3 & 7 \\
  \end{array}
\right)$

The MMS for agent $R$ is~10. The valuation function for $U$ is identical to that for $R$.

In the output of the coarse atomic partial search algorithm, agents $R$ and $C$ either partition among themselves the items of the first and second column, or the items of the first and the third column. In either case, the total value of the atomic bundles is~21, but no combination of atomic bundles has value~10. As all atomic bundles have integer values, either $R$ or $U$ receives a bundle of value at most~9. Hence, coarse atomic partial search does not provide a guarantee better than $\frac{9}{10}.$
\end{proof}

The proof of Theorem~\ref{thm:P+M+M}, that coarse atomic partial search produces a $\rho$-$P^+M^+M$ allocation with $\rho \ge \frac{9}{10}$, follows by running a corresponding MILP that we refer to as MILP9P (P for partial). {MILP9C was solved by an MILP solver (in took less than two seconds to solve MILP9C using https://online-optimizer.appspot.com/), with the result being $\rho=\frac{9}{10}$.} The code for MILP9P, as well as a C++ program that generates MILP9P, can be otained from the authors upon request.

\subsection*{Acknowledgements}

This work was supported by the Israel Science Foundation (grant number 5219/17). The work makes use of the https://online-optimizer.appspot.com/ solver for mixed integer/linear programs (MILPs). Text files containing the code for the MILPs can be obtained from the authors upon request.

\pagebreak

\begin{appendix}


%

\section{MILP9G}
\label{app:milp9b}

Here we present the code for MILP9G, analyzing the allocation algorithm {\em ambitious coarse atomic exhaustive search} of Section~\ref{sec:9Gvariation}.

\footnotesize

\begin{lstlisting}

/* Mixed Integer Linear Program MILP9G.
Analyses the allocation algorithm "ambitious coarse atomic exhaustive search". 
Runs on https://online-optimizer.appspot.com/ (around 70 seconds). 
Three agents R, C, U, additive valuations over goods.
Nine atomic bundles e1 to e9, referred to also as items.
R and C get at least z/(z+1) fraction of maximin share (MMS).
U gets full proportional share (PS).
Strict inequalities for U replaced by weak inequalities with slackness 1/522. */

/* The valuation functions of the three agents */

var e1R>=0;
var e2R>=0;
var e3R>=0;
var e4R>=0;
var e5R>=0;
var e6R>=0;
var e7R>=0;
var e8R>=0;
var e9R>=0;
var e1C>=0;
var e2C>=0;
var e3C>=0;
var e4C>=0;
var e5C>=0;
var e6C>=0;
var e7C>=0;
var e8C>=0;
var e9C>=0;
var e1U>=0;
var e2U>=0;
var e3U>=0;
var e4U>=0;
var e5U>=0;
var e6U>=0;
var e7U>=0;
var e8U>=0;
var e9U>=0;

/* The objective value (the MMS and PS will be z+1) */

var z>=0;

/* Binary variables that will select constraints that need to hold */

var y1>=0, binary;
    .
    .
    .
    .
    .
var y384>=0, binary;

/* The MMS partitions of R and of C, with MMS value z+1 */

subject to c1: e1R + e2R + e3R = z+1;
subject to c2: e4R + e5R + e6R = z+1;
subject to c3: e7R + e8R + e9R = z+1;
subject to c4: e1C + e4C + e7C = z+1;
subject to c5: e2C + e5C + e8C = z+1;
subject to c6: e3C + e6C + e9C = z+1;

/* Row 1 is the unique row that gives C (more than) her MMS and U her PS. */

subject to c7: e4C + e5C + e6C <= z;
subject to c8: e7C + e8C + e9C <= z;
subject to c9: e4U + e5U + e6U <= z+1-1/522;
subject to c10: e7U + e8U + e9U <= z+1-1/522;

/* Column 1 is the unique column that gives R her MMS and U her PS. */

subject to c11: e2R + e5R + e8R <= z;
subject to c12: e3R + e6R + e9R <= z;
subject to c13: e2U + e5U + e8U <= z+1-1/522;
subject to c14: e3U + e6U + e9U <= z+1-1/522;

/* PS of U is z+1 */ 

subject to c15: e1U + e2U + e3U + e4U + e5U + e6U + e7U + e8U +e9U = 3*z+3;

/* minimizing z minimizes z/(z+1) */

minimize statement: z;

/* R gets row 2 (48 partitions for the remaining 6 items) */

/* 6 partitions in which C gets no item from column 3 */

subject to c16: e1C+0<=z+120*(1-y1);
subject to c17: e2U+e3U+e7U+e8U+e9U+0<=z+1-1/522+120*(1-y2);
subject to ycond1: y1+y2>=1;

subject to c18: e2C+0<=z+120*(1-y3);
subject to c19: e1U+e3U+e7U+e8U+e9U+0<=z+1-1/522+120*(1-y4);
subject to ycond2: y3+y4>=1;

subject to c20: e1C+e2C+0<=z+120*(1-y5);
subject to c21: e3U+e7U+e8U+e9U+0<=z+1-1/522+120*(1-y6);
subject to ycond3: y5+y6>=1;

subject to c22: e3C+0<=z+120*(1-y7);
subject to c23: e1U+e2U+e7U+e8U+e9U+0<=z+1-1/522+120*(1-y8);
subject to ycond4: y7+y8>=1;

subject to c24: e1C+e3C+0<=z+120*(1-y9);
subject to c25: e2U+e7U+e8U+e9U+0<=z+1-1/522+120*(1-y10);
subject to ycond5: y9+y10>=1;

subject to c26: e2C+e3C+0<=z+120*(1-y11);
subject to c27: e1U+e7U+e8U+e9U+0<=z+1-1/522+120*(1-y12);
subject to ycond6: y11+y12>=1;

/* 6 partitions in which C gets only e7 from column 3 */

subject to c28: e1C+e7C+0<=z+120*(1-y13);
subject to c29: e2U+e3U+e8U+e9U+0<=z+1-1/522+120*(1-y14);
subject to ycond7: y13+y14>=1;

subject to c30: e2C+e7C+0<=z+120*(1-y15);
subject to c31: e1U+e3U+e8U+e9U+0<=z+1-1/522+120*(1-y16);
subject to ycond8: y15+y16>=1;

subject to c32: e1C+e2C+e7C+0<=z+120*(1-y17);
subject to c33: e3U+e8U+e9U+0<=z+1-1/522+120*(1-y18);
subject to ycond9: y17+y18>=1;

subject to c34: e3C+e7C+0<=z+120*(1-y19);
subject to c35: e1U+e2U+e8U+e9U+0<=z+1-1/522+120*(1-y20);
subject to ycond10: y19+y20>=1;

subject to c36: e1C+e3C+e7C+0<=z+120*(1-y21);
subject to c37: e2U+e8U+e9U+0<=z+1-1/522+120*(1-y22);
subject to ycond11: y21+y22>=1;

subject to c38: e2C+e3C+e7C+0<=z+120*(1-y23);
subject to c39: e1U+e8U+e9U+0<=z+1-1/522+120*(1-y24);
subject to ycond12: y23+y24>=1;

/* 6 partitions in which C gets only e8 from column 3 */

subject to c40: e1C+e8C+0<=z+120*(1-y25);
subject to c41: e2U+e3U+e7U+e9U+0<=z+1-1/522+120*(1-y26);
subject to ycond13: y25+y26>=1;

subject to c42: e2C+e8C+0<=z+120*(1-y27);
subject to c43: e1U+e3U+e7U+e9U+0<=z+1-1/522+120*(1-y28);
subject to ycond14: y27+y28>=1;

subject to c44: e1C+e2C+e8C+0<=z+120*(1-y29);
subject to c45: e3U+e7U+e9U+0<=z+1-1/522+120*(1-y30);
subject to ycond15: y29+y30>=1;

subject to c46: e3C+e8C+0<=z+120*(1-y31);
subject to c47: e1U+e2U+e7U+e9U+0<=z+1-1/522+120*(1-y32);
subject to ycond16: y31+y32>=1;

subject to c48: e1C+e3C+e8C+0<=z+120*(1-y33);
subject to c49: e2U+e7U+e9U+0<=z+1-1/522+120*(1-y34);
subject to ycond17: y33+y34>=1;

subject to c50: e2C+e3C+e8C+0<=z+120*(1-y35);
subject to c51: e1U+e7U+e9U+0<=z+1-1/522+120*(1-y36);
subject to ycond18: y35+y36>=1;

/* 6 partitions in which C gets e7 and e8 from column 3 */

subject to c52: e1C+e7C+e8C+0<=z+120*(1-y37);
subject to c53: e2U+e3U+e9U+0<=z+1-1/522+120*(1-y38);
subject to ycond19: y37+y38>=1;

subject to c54: e2C+e7C+e8C+0<=z+120*(1-y39);
subject to c55: e1U+e3U+e9U+0<=z+1-1/522+120*(1-y40);
subject to ycond20: y39+y40>=1;

subject to c56: e1C+e2C+e7C+e8C+0<=z+120*(1-y41);
subject to c57: e3U+e9U+0<=z+1-1/522+120*(1-y42);
subject to ycond21: y41+y42>=1;

subject to c58: e3C+e7C+e8C+0<=z+120*(1-y43);
subject to c59: e1U+e2U+e9U+0<=z+1-1/522+120*(1-y44);
subject to ycond22: y43+y44>=1;

subject to c60: e1C+e3C+e7C+e8C+0<=z+120*(1-y45);
subject to c61: e2U+e9U+0<=z+1-1/522+120*(1-y46);
subject to ycond23: y45+y46>=1;

subject to c62: e2C+e3C+e7C+e8C+0<=z+120*(1-y47);
subject to c63: e1U+e9U+0<=z+1-1/522+120*(1-y48);
subject to ycond24: y47+y48>=1;

/* 6 partitions in which C gets only e9 from column 3 */

subject to c64: e1C+e9C+0<=z+120*(1-y49);
subject to c65: e2U+e3U+e7U+e8U+0<=z+1-1/522+120*(1-y50);
subject to ycond25: y49+y50>=1;

subject to c66: e2C+e9C+0<=z+120*(1-y51);
subject to c67: e1U+e3U+e7U+e8U+0<=z+1-1/522+120*(1-y52);
subject to ycond26: y51+y52>=1;

subject to c68: e1C+e2C+e9C+0<=z+120*(1-y53);
subject to c69: e3U+e7U+e8U+0<=z+1-1/522+120*(1-y54);
subject to ycond27: y53+y54>=1;

subject to c70: e3C+e9C+0<=z+120*(1-y55);
subject to c71: e1U+e2U+e7U+e8U+0<=z+1-1/522+120*(1-y56);
subject to ycond28: y55+y56>=1;

subject to c72: e1C+e3C+e9C+0<=z+120*(1-y57);
subject to c73: e2U+e7U+e8U+0<=z+1-1/522+120*(1-y58);
subject to ycond29: y57+y58>=1;

subject to c74: e2C+e3C+e9C+0<=z+120*(1-y59);
subject to c75: e1U+e7U+e8U+0<=z+1-1/522+120*(1-y60);
subject to ycond30: y59+y60>=1;

/* 6 partitions in which C gets e7 and e9 from column 3 */

subject to c76: e1C+e7C+e9C+0<=z+120*(1-y61);
subject to c77: e2U+e3U+e8U+0<=z+1-1/522+120*(1-y62);
subject to ycond31: y61+y62>=1;

subject to c78: e2C+e7C+e9C+0<=z+120*(1-y63);
subject to c79: e1U+e3U+e8U+0<=z+1-1/522+120*(1-y64);
subject to ycond32: y63+y64>=1;

subject to c80: e1C+e2C+e7C+e9C+0<=z+120*(1-y65);
subject to c81: e3U+e8U+0<=z+1-1/522+120*(1-y66);
subject to ycond33: y65+y66>=1;

subject to c82: e3C+e7C+e9C+0<=z+120*(1-y67);
subject to c83: e1U+e2U+e8U+0<=z+1-1/522+120*(1-y68);
subject to ycond34: y67+y68>=1;

subject to c84: e1C+e3C+e7C+e9C+0<=z+120*(1-y69);
subject to c85: e2U+e8U+0<=z+1-1/522+120*(1-y70);
subject to ycond35: y69+y70>=1;

subject to c86: e2C+e3C+e7C+e9C+0<=z+120*(1-y71);
subject to c87: e1U+e8U+0<=z+1-1/522+120*(1-y72);
subject to ycond36: y71+y72>=1;

/* 6 partitions in which C gets e8 and e9 from column 3 */

subject to c88: e1C+e8C+e9C+0<=z+120*(1-y73);
subject to c89: e2U+e3U+e7U+0<=z+1-1/522+120*(1-y74);
subject to ycond37: y73+y74>=1;

subject to c90: e2C+e8C+e9C+0<=z+120*(1-y75);
subject to c91: e1U+e3U+e7U+0<=z+1-1/522+120*(1-y76);
subject to ycond38: y75+y76>=1;

subject to c92: e1C+e2C+e8C+e9C+0<=z+120*(1-y77);
subject to c93: e3U+e7U+0<=z+1-1/522+120*(1-y78);
subject to ycond39: y77+y78>=1;

subject to c94: e3C+e8C+e9C+0<=z+120*(1-y79);
subject to c95: e1U+e2U+e7U+0<=z+1-1/522+120*(1-y80);
subject to ycond40: y79+y80>=1;

subject to c96: e1C+e3C+e8C+e9C+0<=z+120*(1-y81);
subject to c97: e2U+e7U+0<=z+1-1/522+120*(1-y82);
subject to ycond41: y81+y82>=1;

subject to c98: e2C+e3C+e8C+e9C+0<=z+120*(1-y83);
subject to c99: e1U+e7U+0<=z+1-1/522+120*(1-y84);
subject to ycond42: y83+y84>=1;

/* 6 partitions in which C gets all of column 3 */

subject to c100: e1C+e7C+e8C+e9C+0<=z+120*(1-y85);
subject to c101: e2U+e3U+0<=z+1-1/522+120*(1-y86);
subject to ycond43: y85+y86>=1;

subject to c102: e2C+e7C+e8C+e9C+0<=z+120*(1-y87);
subject to c103: e1U+e3U+0<=z+1-1/522+120*(1-y88);
subject to ycond44: y87+y88>=1;

subject to c104: e1C+e2C+e7C+e8C+e9C+0<=z+120*(1-y89);
subject to c105: e3U+0<=z+1-1/522+120*(1-y90);
subject to ycond45: y89+y90>=1;

subject to c106: e3C+e7C+e8C+e9C+0<=z+120*(1-y91);
subject to c107: e1U+e2U+0<=z+1-1/522+120*(1-y92);
subject to ycond46: y91+y92>=1;

subject to c108: e1C+e3C+e7C+e8C+e9C+0<=z+120*(1-y93);
subject to c109: e2U+0<=z+1-1/522+120*(1-y94);
subject to ycond47: y93+y94>=1;

subject to c110: e2C+e3C+e7C+e8C+e9C+0<=z+120*(1-y95);
subject to c111: e1U+0<=z+1-1/522+120*(1-y96);
subject to ycond48: y95+y96>=1;

/* R gets row 3 (48 partitions for the remaining 6 items) */

subject to c112: e1C+0<=z+120*(1-y97);
subject to c113: e2U+e3U+e4U+e5U+e6U+0<=z+1-1/522+120*(1-y98);
subject to ycond49: y97+y98>=1;

subject to c114: e2C+0<=z+120*(1-y99);
subject to c115: e1U+e3U+e4U+e5U+e6U+0<=z+1-1/522+120*(1-y100);
subject to ycond50: y99+y100>=1;

subject to c116: e1C+e2C+0<=z+120*(1-y101);
subject to c117: e3U+e4U+e5U+e6U+0<=z+1-1/522+120*(1-y102);
subject to ycond51: y101+y102>=1;

subject to c118: e3C+0<=z+120*(1-y103);
subject to c119: e1U+e2U+e4U+e5U+e6U+0<=z+1-1/522+120*(1-y104);
subject to ycond52: y103+y104>=1;

subject to c120: e1C+e3C+0<=z+120*(1-y105);
subject to c121: e2U+e4U+e5U+e6U+0<=z+1-1/522+120*(1-y106);
subject to ycond53: y105+y106>=1;

subject to c122: e2C+e3C+0<=z+120*(1-y107);
subject to c123: e1U+e4U+e5U+e6U+0<=z+1-1/522+120*(1-y108);
subject to ycond54: y107+y108>=1;

subject to c124: e1C+e4C+0<=z+120*(1-y109);
subject to c125: e2U+e3U+e5U+e6U+0<=z+1-1/522+120*(1-y110);
subject to ycond55: y109+y110>=1;

subject to c126: e2C+e4C+0<=z+120*(1-y111);
subject to c127: e1U+e3U+e5U+e6U+0<=z+1-1/522+120*(1-y112);
subject to ycond56: y111+y112>=1;

subject to c128: e1C+e2C+e4C+0<=z+120*(1-y113);
subject to c129: e3U+e5U+e6U+0<=z+1-1/522+120*(1-y114);
subject to ycond57: y113+y114>=1;

subject to c130: e3C+e4C+0<=z+120*(1-y115);
subject to c131: e1U+e2U+e5U+e6U+0<=z+1-1/522+120*(1-y116);
subject to ycond58: y115+y116>=1;

subject to c132: e1C+e3C+e4C+0<=z+120*(1-y117);
subject to c133: e2U+e5U+e6U+0<=z+1-1/522+120*(1-y118);
subject to ycond59: y117+y118>=1;

subject to c134: e2C+e3C+e4C+0<=z+120*(1-y119);
subject to c135: e1U+e5U+e6U+0<=z+1-1/522+120*(1-y120);
subject to ycond60: y119+y120>=1;

subject to c136: e1C+e5C+0<=z+120*(1-y121);
subject to c137: e2U+e3U+e4U+e6U+0<=z+1-1/522+120*(1-y122);
subject to ycond61: y121+y122>=1;

subject to c138: e2C+e5C+0<=z+120*(1-y123);
subject to c139: e1U+e3U+e4U+e6U+0<=z+1-1/522+120*(1-y124);
subject to ycond62: y123+y124>=1;

subject to c140: e1C+e2C+e5C+0<=z+120*(1-y125);
subject to c141: e3U+e4U+e6U+0<=z+1-1/522+120*(1-y126);
subject to ycond63: y125+y126>=1;

subject to c142: e3C+e5C+0<=z+120*(1-y127);
subject to c143: e1U+e2U+e4U+e6U+0<=z+1-1/522+120*(1-y128);
subject to ycond64: y127+y128>=1;

subject to c144: e1C+e3C+e5C+0<=z+120*(1-y129);
subject to c145: e2U+e4U+e6U+0<=z+1-1/522+120*(1-y130);
subject to ycond65: y129+y130>=1;

subject to c146: e2C+e3C+e5C+0<=z+120*(1-y131);
subject to c147: e1U+e4U+e6U+0<=z+1-1/522+120*(1-y132);
subject to ycond66: y131+y132>=1;

subject to c148: e1C+e4C+e5C+0<=z+120*(1-y133);
subject to c149: e2U+e3U+e6U+0<=z+1-1/522+120*(1-y134);
subject to ycond67: y133+y134>=1;

subject to c150: e2C+e4C+e5C+0<=z+120*(1-y135);
subject to c151: e1U+e3U+e6U+0<=z+1-1/522+120*(1-y136);
subject to ycond68: y135+y136>=1;

subject to c152: e1C+e2C+e4C+e5C+0<=z+120*(1-y137);
subject to c153: e3U+e6U+0<=z+1-1/522+120*(1-y138);
subject to ycond69: y137+y138>=1;

subject to c154: e3C+e4C+e5C+0<=z+120*(1-y139);
subject to c155: e1U+e2U+e6U+0<=z+1-1/522+120*(1-y140);
subject to ycond70: y139+y140>=1;

subject to c156: e1C+e3C+e4C+e5C+0<=z+120*(1-y141);
subject to c157: e2U+e6U+0<=z+1-1/522+120*(1-y142);
subject to ycond71: y141+y142>=1;

subject to c158: e2C+e3C+e4C+e5C+0<=z+120*(1-y143);
subject to c159: e1U+e6U+0<=z+1-1/522+120*(1-y144);
subject to ycond72: y143+y144>=1;

subject to c160: e1C+e6C+0<=z+120*(1-y145);
subject to c161: e2U+e3U+e4U+e5U+0<=z+1-1/522+120*(1-y146);
subject to ycond73: y145+y146>=1;

subject to c162: e2C+e6C+0<=z+120*(1-y147);
subject to c163: e1U+e3U+e4U+e5U+0<=z+1-1/522+120*(1-y148);
subject to ycond74: y147+y148>=1;

subject to c164: e1C+e2C+e6C+0<=z+120*(1-y149);
subject to c165: e3U+e4U+e5U+0<=z+1-1/522+120*(1-y150);
subject to ycond75: y149+y150>=1;

subject to c166: e3C+e6C+0<=z+120*(1-y151);
subject to c167: e1U+e2U+e4U+e5U+0<=z+1-1/522+120*(1-y152);
subject to ycond76: y151+y152>=1;

subject to c168: e1C+e3C+e6C+0<=z+120*(1-y153);
subject to c169: e2U+e4U+e5U+0<=z+1-1/522+120*(1-y154);
subject to ycond77: y153+y154>=1;

subject to c170: e2C+e3C+e6C+0<=z+120*(1-y155);
subject to c171: e1U+e4U+e5U+0<=z+1-1/522+120*(1-y156);
subject to ycond78: y155+y156>=1;

subject to c172: e1C+e4C+e6C+0<=z+120*(1-y157);
subject to c173: e2U+e3U+e5U+0<=z+1-1/522+120*(1-y158);
subject to ycond79: y157+y158>=1;

subject to c174: e2C+e4C+e6C+0<=z+120*(1-y159);
subject to c175: e1U+e3U+e5U+0<=z+1-1/522+120*(1-y160);
subject to ycond80: y159+y160>=1;

subject to c176: e1C+e2C+e4C+e6C+0<=z+120*(1-y161);
subject to c177: e3U+e5U+0<=z+1-1/522+120*(1-y162);
subject to ycond81: y161+y162>=1;

subject to c178: e3C+e4C+e6C+0<=z+120*(1-y163);
subject to c179: e1U+e2U+e5U+0<=z+1-1/522+120*(1-y164);
subject to ycond82: y163+y164>=1;

subject to c180: e1C+e3C+e4C+e6C+0<=z+120*(1-y165);
subject to c181: e2U+e5U+0<=z+1-1/522+120*(1-y166);
subject to ycond83: y165+y166>=1;

subject to c182: e2C+e3C+e4C+e6C+0<=z+120*(1-y167);
subject to c183: e1U+e5U+0<=z+1-1/522+120*(1-y168);
subject to ycond84: y167+y168>=1;

subject to c184: e1C+e5C+e6C+0<=z+120*(1-y169);
subject to c185: e2U+e3U+e4U+0<=z+1-1/522+120*(1-y170);
subject to ycond85: y169+y170>=1;

subject to c186: e2C+e5C+e6C+0<=z+120*(1-y171);
subject to c187: e1U+e3U+e4U+0<=z+1-1/522+120*(1-y172);
subject to ycond86: y171+y172>=1;

subject to c188: e1C+e2C+e5C+e6C+0<=z+120*(1-y173);
subject to c189: e3U+e4U+0<=z+1-1/522+120*(1-y174);
subject to ycond87: y173+y174>=1;

subject to c190: e3C+e5C+e6C+0<=z+120*(1-y175);
subject to c191: e1U+e2U+e4U+0<=z+1-1/522+120*(1-y176);
subject to ycond88: y175+y176>=1;

subject to c192: e1C+e3C+e5C+e6C+0<=z+120*(1-y177);
subject to c193: e2U+e4U+0<=z+1-1/522+120*(1-y178);
subject to ycond89: y177+y178>=1;

subject to c194: e2C+e3C+e5C+e6C+0<=z+120*(1-y179);
subject to c195: e1U+e4U+0<=z+1-1/522+120*(1-y180);
subject to ycond90: y179+y180>=1;

subject to c196: e1C+e4C+e5C+e6C+0<=z+120*(1-y181);
subject to c197: e2U+e3U+0<=z+1-1/522+120*(1-y182);
subject to ycond91: y181+y182>=1;

subject to c198: e2C+e4C+e5C+e6C+0<=z+120*(1-y183);
subject to c199: e1U+e3U+0<=z+1-1/522+120*(1-y184);
subject to ycond92: y183+y184>=1;

subject to c200: e1C+e2C+e4C+e5C+e6C+0<=z+120*(1-y185);
subject to c201: e3U+0<=z+1-1/522+120*(1-y186);
subject to ycond93: y185+y186>=1;

subject to c202: e3C+e4C+e5C+e6C+0<=z+120*(1-y187);
subject to c203: e1U+e2U+0<=z+1-1/522+120*(1-y188);
subject to ycond94: y187+y188>=1;

subject to c204: e1C+e3C+e4C+e5C+e6C+0<=z+120*(1-y189);
subject to c205: e2U+0<=z+1-1/522+120*(1-y190);
subject to ycond95: y189+y190>=1;

subject to c206: e2C+e3C+e4C+e5C+e6C+0<=z+120*(1-y191);
subject to c207: e1U+0<=z+1-1/522+120*(1-y192);
subject to ycond96: y191+y192>=1;

/* C gets column 2 (48 partitions for the remaining 6 items) */

subject to c208: e1R+0<=z+120*(1-y193);
subject to c209: e3U+e4U+e6U+e7U+e9U+0<=z+1-1/522+120*(1-y194);
subject to ycond97: y193+y194>=1;

subject to c210: e1R+e3R+0<=z+120*(1-y195);
subject to c211: e4U+e6U+e7U+e9U+0<=z+1-1/522+120*(1-y196);
subject to ycond98: y195+y196>=1;

subject to c212: e4R+0<=z+120*(1-y197);
subject to c213: e1U+e3U+e6U+e7U+e9U+0<=z+1-1/522+120*(1-y198);
subject to ycond99: y197+y198>=1;

subject to c214: e1R+e4R+0<=z+120*(1-y199);
subject to c215: e3U+e6U+e7U+e9U+0<=z+1-1/522+120*(1-y200);
subject to ycond100: y199+y200>=1;

subject to c216: e3R+e4R+0<=z+120*(1-y201);
subject to c217: e1U+e6U+e7U+e9U+0<=z+1-1/522+120*(1-y202);
subject to ycond101: y201+y202>=1;

subject to c218: e1R+e3R+e4R+0<=z+120*(1-y203);
subject to c219: e6U+e7U+e9U+0<=z+1-1/522+120*(1-y204);
subject to ycond102: y203+y204>=1;

subject to c220: e1R+e6R+0<=z+120*(1-y205);
subject to c221: e3U+e4U+e7U+e9U+0<=z+1-1/522+120*(1-y206);
subject to ycond103: y205+y206>=1;

subject to c222: e1R+e3R+e6R+0<=z+120*(1-y207);
subject to c223: e4U+e7U+e9U+0<=z+1-1/522+120*(1-y208);
subject to ycond104: y207+y208>=1;

subject to c224: e4R+e6R+0<=z+120*(1-y209);
subject to c225: e1U+e3U+e7U+e9U+0<=z+1-1/522+120*(1-y210);
subject to ycond105: y209+y210>=1;

subject to c226: e1R+e4R+e6R+0<=z+120*(1-y211);
subject to c227: e3U+e7U+e9U+0<=z+1-1/522+120*(1-y212);
subject to ycond106: y211+y212>=1;

subject to c228: e3R+e4R+e6R+0<=z+120*(1-y213);
subject to c229: e1U+e7U+e9U+0<=z+1-1/522+120*(1-y214);
subject to ycond107: y213+y214>=1;

subject to c230: e1R+e3R+e4R+e6R+0<=z+120*(1-y215);
subject to c231: e7U+e9U+0<=z+1-1/522+120*(1-y216);
subject to ycond108: y215+y216>=1;

subject to c232: e7R+0<=z+120*(1-y217);
subject to c233: e1U+e3U+e4U+e6U+e9U+0<=z+1-1/522+120*(1-y218);
subject to ycond109: y217+y218>=1;

subject to c234: e1R+e7R+0<=z+120*(1-y219);
subject to c235: e3U+e4U+e6U+e9U+0<=z+1-1/522+120*(1-y220);
subject to ycond110: y219+y220>=1;

subject to c236: e3R+e7R+0<=z+120*(1-y221);
subject to c237: e1U+e4U+e6U+e9U+0<=z+1-1/522+120*(1-y222);
subject to ycond111: y221+y222>=1;

subject to c238: e1R+e3R+e7R+0<=z+120*(1-y223);
subject to c239: e4U+e6U+e9U+0<=z+1-1/522+120*(1-y224);
subject to ycond112: y223+y224>=1;

subject to c240: e4R+e7R+0<=z+120*(1-y225);
subject to c241: e1U+e3U+e6U+e9U+0<=z+1-1/522+120*(1-y226);
subject to ycond113: y225+y226>=1;

subject to c242: e3R+e4R+e7R+0<=z+120*(1-y227);
subject to c243: e1U+e6U+e9U+0<=z+1-1/522+120*(1-y228);
subject to ycond114: y227+y228>=1;

subject to c244: e6R+e7R+0<=z+120*(1-y229);
subject to c245: e1U+e3U+e4U+e9U+0<=z+1-1/522+120*(1-y230);
subject to ycond115: y229+y230>=1;

subject to c246: e1R+e6R+e7R+0<=z+120*(1-y231);
subject to c247: e3U+e4U+e9U+0<=z+1-1/522+120*(1-y232);
subject to ycond116: y231+y232>=1;

subject to c248: e3R+e6R+e7R+0<=z+120*(1-y233);
subject to c249: e1U+e4U+e9U+0<=z+1-1/522+120*(1-y234);
subject to ycond117: y233+y234>=1;

subject to c250: e1R+e3R+e6R+e7R+0<=z+120*(1-y235);
subject to c251: e4U+e9U+0<=z+1-1/522+120*(1-y236);
subject to ycond118: y235+y236>=1;

subject to c252: e4R+e6R+e7R+0<=z+120*(1-y237);
subject to c253: e1U+e3U+e9U+0<=z+1-1/522+120*(1-y238);
subject to ycond119: y237+y238>=1;

subject to c254: e3R+e4R+e6R+e7R+0<=z+120*(1-y239);
subject to c255: e1U+e9U+0<=z+1-1/522+120*(1-y240);
subject to ycond120: y239+y240>=1;

subject to c256: e1R+e9R+0<=z+120*(1-y241);
subject to c257: e3U+e4U+e6U+e7U+0<=z+1-1/522+120*(1-y242);
subject to ycond121: y241+y242>=1;

subject to c258: e1R+e3R+e9R+0<=z+120*(1-y243);
subject to c259: e4U+e6U+e7U+0<=z+1-1/522+120*(1-y244);
subject to ycond122: y243+y244>=1;

subject to c260: e4R+e9R+0<=z+120*(1-y245);
subject to c261: e1U+e3U+e6U+e7U+0<=z+1-1/522+120*(1-y246);
subject to ycond123: y245+y246>=1;

subject to c262: e1R+e4R+e9R+0<=z+120*(1-y247);
subject to c263: e3U+e6U+e7U+0<=z+1-1/522+120*(1-y248);
subject to ycond124: y247+y248>=1;

subject to c264: e3R+e4R+e9R+0<=z+120*(1-y249);
subject to c265: e1U+e6U+e7U+0<=z+1-1/522+120*(1-y250);
subject to ycond125: y249+y250>=1;

subject to c266: e1R+e3R+e4R+e9R+0<=z+120*(1-y251);
subject to c267: e6U+e7U+0<=z+1-1/522+120*(1-y252);
subject to ycond126: y251+y252>=1;

subject to c268: e1R+e6R+e9R+0<=z+120*(1-y253);
subject to c269: e3U+e4U+e7U+0<=z+1-1/522+120*(1-y254);
subject to ycond127: y253+y254>=1;

subject to c270: e1R+e3R+e6R+e9R+0<=z+120*(1-y255);
subject to c271: e4U+e7U+0<=z+1-1/522+120*(1-y256);
subject to ycond128: y255+y256>=1;

subject to c272: e4R+e6R+e9R+0<=z+120*(1-y257);
subject to c273: e1U+e3U+e7U+0<=z+1-1/522+120*(1-y258);
subject to ycond129: y257+y258>=1;

subject to c274: e1R+e4R+e6R+e9R+0<=z+120*(1-y259);
subject to c275: e3U+e7U+0<=z+1-1/522+120*(1-y260);
subject to ycond130: y259+y260>=1;

subject to c276: e3R+e4R+e6R+e9R+0<=z+120*(1-y261);
subject to c277: e1U+e7U+0<=z+1-1/522+120*(1-y262);
subject to ycond131: y261+y262>=1;

subject to c278: e1R+e3R+e4R+e6R+e9R+0<=z+120*(1-y263);
subject to c279: e7U+0<=z+1-1/522+120*(1-y264);
subject to ycond132: y263+y264>=1;

subject to c280: e7R+e9R+0<=z+120*(1-y265);
subject to c281: e1U+e3U+e4U+e6U+0<=z+1-1/522+120*(1-y266);
subject to ycond133: y265+y266>=1;

subject to c282: e1R+e7R+e9R+0<=z+120*(1-y267);
subject to c283: e3U+e4U+e6U+0<=z+1-1/522+120*(1-y268);
subject to ycond134: y267+y268>=1;

subject to c284: e3R+e7R+e9R+0<=z+120*(1-y269);
subject to c285: e1U+e4U+e6U+0<=z+1-1/522+120*(1-y270);
subject to ycond135: y269+y270>=1;

subject to c286: e1R+e3R+e7R+e9R+0<=z+120*(1-y271);
subject to c287: e4U+e6U+0<=z+1-1/522+120*(1-y272);
subject to ycond136: y271+y272>=1;

subject to c288: e4R+e7R+e9R+0<=z+120*(1-y273);
subject to c289: e1U+e3U+e6U+0<=z+1-1/522+120*(1-y274);
subject to ycond137: y273+y274>=1;

subject to c290: e3R+e4R+e7R+e9R+0<=z+120*(1-y275);
subject to c291: e1U+e6U+0<=z+1-1/522+120*(1-y276);
subject to ycond138: y275+y276>=1;

subject to c292: e6R+e7R+e9R+0<=z+120*(1-y277);
subject to c293: e1U+e3U+e4U+0<=z+1-1/522+120*(1-y278);
subject to ycond139: y277+y278>=1;

subject to c294: e1R+e6R+e7R+e9R+0<=z+120*(1-y279);
subject to c295: e3U+e4U+0<=z+1-1/522+120*(1-y280);
subject to ycond140: y279+y280>=1;

subject to c296: e3R+e6R+e7R+e9R+0<=z+120*(1-y281);
subject to c297: e1U+e4U+0<=z+1-1/522+120*(1-y282);
subject to ycond141: y281+y282>=1;

subject to c298: e1R+e3R+e6R+e7R+e9R+0<=z+120*(1-y283);
subject to c299: e4U+0<=z+1-1/522+120*(1-y284);
subject to ycond142: y283+y284>=1;

subject to c300: e4R+e6R+e7R+e9R+0<=z+120*(1-y285);
subject to c301: e1U+e3U+0<=z+1-1/522+120*(1-y286);
subject to ycond143: y285+y286>=1;

subject to c302: e3R+e4R+e6R+e7R+e9R+0<=z+120*(1-y287);
subject to c303: e1U+0<=z+1-1/522+120*(1-y288);
subject to ycond144: y287+y288>=1;

/* C gets column 3 (48 partitions for the remaining 6 items) */

subject to c304: e1R+0<=z+120*(1-y289);
subject to c305: e2U+e4U+e5U+e7U+e8U+0<=z+1-1/522+120*(1-y290);
subject to ycond145: y289+y290>=1;

subject to c306: e1R+e2R+0<=z+120*(1-y291);
subject to c307: e4U+e5U+e7U+e8U+0<=z+1-1/522+120*(1-y292);
subject to ycond146: y291+y292>=1;

subject to c308: e4R+0<=z+120*(1-y293);
subject to c309: e1U+e2U+e5U+e7U+e8U+0<=z+1-1/522+120*(1-y294);
subject to ycond147: y293+y294>=1;

subject to c310: e1R+e4R+0<=z+120*(1-y295);
subject to c311: e2U+e5U+e7U+e8U+0<=z+1-1/522+120*(1-y296);
subject to ycond148: y295+y296>=1;

subject to c312: e2R+e4R+0<=z+120*(1-y297);
subject to c313: e1U+e5U+e7U+e8U+0<=z+1-1/522+120*(1-y298);
subject to ycond149: y297+y298>=1;

subject to c314: e1R+e2R+e4R+0<=z+120*(1-y299);
subject to c315: e5U+e7U+e8U+0<=z+1-1/522+120*(1-y300);
subject to ycond150: y299+y300>=1;

subject to c316: e1R+e5R+0<=z+120*(1-y301);
subject to c317: e2U+e4U+e7U+e8U+0<=z+1-1/522+120*(1-y302);
subject to ycond151: y301+y302>=1;

subject to c318: e1R+e2R+e5R+0<=z+120*(1-y303);
subject to c319: e4U+e7U+e8U+0<=z+1-1/522+120*(1-y304);
subject to ycond152: y303+y304>=1;

subject to c320: e4R+e5R+0<=z+120*(1-y305);
subject to c321: e1U+e2U+e7U+e8U+0<=z+1-1/522+120*(1-y306);
subject to ycond153: y305+y306>=1;

subject to c322: e1R+e4R+e5R+0<=z+120*(1-y307);
subject to c323: e2U+e7U+e8U+0<=z+1-1/522+120*(1-y308);
subject to ycond154: y307+y308>=1;

subject to c324: e2R+e4R+e5R+0<=z+120*(1-y309);
subject to c325: e1U+e7U+e8U+0<=z+1-1/522+120*(1-y310);
subject to ycond155: y309+y310>=1;

subject to c326: e1R+e2R+e4R+e5R+0<=z+120*(1-y311);
subject to c327: e7U+e8U+0<=z+1-1/522+120*(1-y312);
subject to ycond156: y311+y312>=1;

subject to c328: e7R+0<=z+120*(1-y313);
subject to c329: e1U+e2U+e4U+e5U+e8U+0<=z+1-1/522+120*(1-y314);
subject to ycond157: y313+y314>=1;

subject to c330: e1R+e7R+0<=z+120*(1-y315);
subject to c331: e2U+e4U+e5U+e8U+0<=z+1-1/522+120*(1-y316);
subject to ycond158: y315+y316>=1;

subject to c332: e2R+e7R+0<=z+120*(1-y317);
subject to c333: e1U+e4U+e5U+e8U+0<=z+1-1/522+120*(1-y318);
subject to ycond159: y317+y318>=1;

subject to c334: e1R+e2R+e7R+0<=z+120*(1-y319);
subject to c335: e4U+e5U+e8U+0<=z+1-1/522+120*(1-y320);
subject to ycond160: y319+y320>=1;

subject to c336: e4R+e7R+0<=z+120*(1-y321);
subject to c337: e1U+e2U+e5U+e8U+0<=z+1-1/522+120*(1-y322);
subject to ycond161: y321+y322>=1;

subject to c338: e2R+e4R+e7R+0<=z+120*(1-y323);
subject to c339: e1U+e5U+e8U+0<=z+1-1/522+120*(1-y324);
subject to ycond162: y323+y324>=1;

subject to c340: e5R+e7R+0<=z+120*(1-y325);
subject to c341: e1U+e2U+e4U+e8U+0<=z+1-1/522+120*(1-y326);
subject to ycond163: y325+y326>=1;

subject to c342: e1R+e5R+e7R+0<=z+120*(1-y327);
subject to c343: e2U+e4U+e8U+0<=z+1-1/522+120*(1-y328);
subject to ycond164: y327+y328>=1;

subject to c344: e2R+e5R+e7R+0<=z+120*(1-y329);
subject to c345: e1U+e4U+e8U+0<=z+1-1/522+120*(1-y330);
subject to ycond165: y329+y330>=1;

subject to c346: e1R+e2R+e5R+e7R+0<=z+120*(1-y331);
subject to c347: e4U+e8U+0<=z+1-1/522+120*(1-y332);
subject to ycond166: y331+y332>=1;

subject to c348: e4R+e5R+e7R+0<=z+120*(1-y333);
subject to c349: e1U+e2U+e8U+0<=z+1-1/522+120*(1-y334);
subject to ycond167: y333+y334>=1;

subject to c350: e2R+e4R+e5R+e7R+0<=z+120*(1-y335);
subject to c351: e1U+e8U+0<=z+1-1/522+120*(1-y336);
subject to ycond168: y335+y336>=1;

subject to c352: e1R+e8R+0<=z+120*(1-y337);
subject to c353: e2U+e4U+e5U+e7U+0<=z+1-1/522+120*(1-y338);
subject to ycond169: y337+y338>=1;

subject to c354: e1R+e2R+e8R+0<=z+120*(1-y339);
subject to c355: e4U+e5U+e7U+0<=z+1-1/522+120*(1-y340);
subject to ycond170: y339+y340>=1;

subject to c356: e4R+e8R+0<=z+120*(1-y341);
subject to c357: e1U+e2U+e5U+e7U+0<=z+1-1/522+120*(1-y342);
subject to ycond171: y341+y342>=1;

subject to c358: e1R+e4R+e8R+0<=z+120*(1-y343);
subject to c359: e2U+e5U+e7U+0<=z+1-1/522+120*(1-y344);
subject to ycond172: y343+y344>=1;

subject to c360: e2R+e4R+e8R+0<=z+120*(1-y345);
subject to c361: e1U+e5U+e7U+0<=z+1-1/522+120*(1-y346);
subject to ycond173: y345+y346>=1;

subject to c362: e1R+e2R+e4R+e8R+0<=z+120*(1-y347);
subject to c363: e5U+e7U+0<=z+1-1/522+120*(1-y348);
subject to ycond174: y347+y348>=1;

subject to c364: e1R+e5R+e8R+0<=z+120*(1-y349);
subject to c365: e2U+e4U+e7U+0<=z+1-1/522+120*(1-y350);
subject to ycond175: y349+y350>=1;

subject to c366: e1R+e2R+e5R+e8R+0<=z+120*(1-y351);
subject to c367: e4U+e7U+0<=z+1-1/522+120*(1-y352);
subject to ycond176: y351+y352>=1;

subject to c368: e4R+e5R+e8R+0<=z+120*(1-y353);
subject to c369: e1U+e2U+e7U+0<=z+1-1/522+120*(1-y354);
subject to ycond177: y353+y354>=1;

subject to c370: e1R+e4R+e5R+e8R+0<=z+120*(1-y355);
subject to c371: e2U+e7U+0<=z+1-1/522+120*(1-y356);
subject to ycond178: y355+y356>=1;

subject to c372: e2R+e4R+e5R+e8R+0<=z+120*(1-y357);
subject to c373: e1U+e7U+0<=z+1-1/522+120*(1-y358);
subject to ycond179: y357+y358>=1;

subject to c374: e1R+e2R+e4R+e5R+e8R+0<=z+120*(1-y359);
subject to c375: e7U+0<=z+1-1/522+120*(1-y360);
subject to ycond180: y359+y360>=1;

subject to c376: e7R+e8R+0<=z+120*(1-y361);
subject to c377: e1U+e2U+e4U+e5U+0<=z+1-1/522+120*(1-y362);
subject to ycond181: y361+y362>=1;

subject to c378: e1R+e7R+e8R+0<=z+120*(1-y363);
subject to c379: e2U+e4U+e5U+0<=z+1-1/522+120*(1-y364);
subject to ycond182: y363+y364>=1;

subject to c380: e2R+e7R+e8R+0<=z+120*(1-y365);
subject to c381: e1U+e4U+e5U+0<=z+1-1/522+120*(1-y366);
subject to ycond183: y365+y366>=1;

subject to c382: e1R+e2R+e7R+e8R+0<=z+120*(1-y367);
subject to c383: e4U+e5U+0<=z+1-1/522+120*(1-y368);
subject to ycond184: y367+y368>=1;

subject to c384: e4R+e7R+e8R+0<=z+120*(1-y369);
subject to c385: e1U+e2U+e5U+0<=z+1-1/522+120*(1-y370);
subject to ycond185: y369+y370>=1;

subject to c386: e2R+e4R+e7R+e8R+0<=z+120*(1-y371);
subject to c387: e1U+e5U+0<=z+1-1/522+120*(1-y372);
subject to ycond186: y371+y372>=1;

subject to c388: e5R+e7R+e8R+0<=z+120*(1-y373);
subject to c389: e1U+e2U+e4U+0<=z+1-1/522+120*(1-y374);
subject to ycond187: y373+y374>=1;

subject to c390: e1R+e5R+e7R+e8R+0<=z+120*(1-y375);
subject to c391: e2U+e4U+0<=z+1-1/522+120*(1-y376);
subject to ycond188: y375+y376>=1;

subject to c392: e2R+e5R+e7R+e8R+0<=z+120*(1-y377);
subject to c393: e1U+e4U+0<=z+1-1/522+120*(1-y378);
subject to ycond189: y377+y378>=1;

subject to c394: e1R+e2R+e5R+e7R+e8R+0<=z+120*(1-y379);
subject to c395: e4U+0<=z+1-1/522+120*(1-y380);
subject to ycond190: y379+y380>=1;

subject to c396: e4R+e5R+e7R+e8R+0<=z+120*(1-y381);
subject to c397: e1U+e2U+0<=z+1-1/522+120*(1-y382);
subject to ycond191: y381+y382>=1;

subject to c398: e2R+e4R+e5R+e7R+e8R+0<=z+120*(1-y383);
subject to c399: e1U+0<=z+1-1/522+120*(1-y384);
subject to ycond192: y383+y384>=1;

end;
\end{lstlisting}



\pagebreak

\section{A C++ program that generates MILP9G}
\label{app:9bprogramGoods}

\normalsize

As it is tedious to explicitly write all constraints of our MILPs, the MILPs where generated automatically by C++ programs. We present here the C++ program that we used in order to generate the constraints of MILP9G.

\scriptsize

\begin{lstlisting}
#include <iostream>
#include <cmath>
#include <fstream>
#include <cstring>
#include <algorithm>
#include <bits/stdc++.h>
#include <numeric>

using namespace std;

int cond=18;
int ycond=1;
vector<string> Subset;

void AllSubsets(string *arr, int n)
{
    int card=pow(2, n);
    for(int i=0; i<card; i++)
    {
        string t="";
        for(int j=0; j<n; j++)
        {
            if((i&(1<<j))!=0)
            {
                t+=arr[j];
            }
        }
        Subset.push_back(t);
    }
}

bool Reoccurrence(string arr)
{
    for(int i=0; i<arr.length(); i++)
    {
        if(i!=arr.length()-1)
            for(int j=i+1; j<arr.length(); j++)
            {
                if(arr[j]==arr[i])
                    return 1; //there is a repeated symbol
            }
    }
    return 0;//no repeated symbols

}

bool BundleRule(string strR, string strC, int selection)//Selection of Bundles
{
    if(selection==0)//Agent R gets r2
    {
        //123
        //456
        //789
        if(strR!="456")
            return 0;
        else
        {
            int counter=0;
            if(strC.find("1")!=-1)
                counter++;
            if(strC.find("2")!=-1)
                counter++;
            if(strC.find("3")!=-1)
                counter++;
            if(counter>0&&counter<3)
                return 1;
            return 0;
        }
    }
    else if(selection==1)//Agent R gets r3
    {
        if(strR!="789")
            return 0;
        else
        {
            int counter=0;
            if(strC.find("1")!=-1)
                counter++;
            if(strC.find("2")!=-1)
                counter++;
            if(strC.find("3")!=-1)
                counter++;
            if(counter>0&&counter<3)
                return 1;
            return 0;
        }
    }
    else if(selection==2)//Agent C gets c2
    {
        if(strC!="258")
            return 0;
        else
        {
            int counter=0;
            if(strR.find("1")!=-1)
                counter++;
            if(strR.find("4")!=-1)
                counter++;
            if(strR.find("7")!=-1)
                counter++;
            if(counter>0&&counter<3)
                return 1;
            return 0;
        }
    }
    else if(selection==3)//Agent C gets c3
    {
        if(strC!="369")
            return 0;
        else
        {
            int counter=0;
            if(strR.find("1")!=-1)
                counter++;
            if(strR.find("4")!=-1)
                counter++;
            if(strR.find("7")!=-1)
                counter++;
            if(counter>0&&counter<3)
                return 1;
            return 0;
        }
    }
    else
        return 0;
}

int main()
{
    string items[9]={"1", "2", "3", "4", "5", "6", "7", "8", "9"};
    AllSubsets(items, 9);
    vector<string> Rbundle;
    vector<string> Cbundle;
    for(int i=0;i<512; i++)
    {
        string y=Subset[i];
        if(y!=""&&y!="123456789")
        {
            Rbundle.push_back(y);
            Cbundle.push_back(y);
        }
    }
    ofstream outprogram ("MILP9G.txt");
    for(int i=0;i<9;i++)
        outprogram<<"var e"<<i+1<<"R>=0;"<<endl;
    for(int i=0;i<9;i++)
        outprogram<<"var e"<<i+1<<"C>=0;"<<endl;
    for(int i=0;i<9;i++)
        outprogram<<"var e"<<i+1<<"U>=0;"<<endl;
    outprogram<<endl;
    outprogram<<"var z>=0;"<<endl;
    outprogram<<endl;
    for(int y=0;y<384;y++)
        outprogram<<"var y"<<y+1<<">=0, binary;"<<endl;
    outprogram<<endl;
    outprogram<<"subject to c1: e1R + e2R + e3R = z+1;"<<endl;
    outprogram<<"subject to c2: e4R + e5R + e6R = z+1;"<<endl;
    outprogram<<"subject to c3: e7R + e8R + e9R = z+1;"<<endl;
    outprogram<<"subject to c4: e1C + e4C + e7C = z+1;"<<endl;
    outprogram<<"subject to c5: e2C + e5C + e8C = z+1;"<<endl;
    outprogram<<"subject to c6: e3C + e6C + e9C = z+1;"<<endl;
    outprogram<<endl;

    outprogram<<"subject to c7: e2R + e5R + e8R <= z;"<<endl;
    outprogram<<"subject to c8: e3R + e6R + e9R <= z;"<<endl;
    outprogram<<endl;

    outprogram<<"subject to c9: e4C + e5C + e6C <= z;"<<endl;
    outprogram<<"subject to c10: e7C + e8C + e9C <= z;"<<endl;
    outprogram<<endl;

    outprogram<<"subject to c11: e4U + e5U + e6U <= z+1-1/522;"<<endl;
    outprogram<<"subject to c12: e7U + e8U + e9U <= z+1-1/522;"<<endl;
    outprogram<<endl;

    outprogram<<"subject to c13: e2R + e5R + e8R <= z;"<<endl;
    outprogram<<"subject to c14: e3R + e6R + e9R <= z;"<<endl;
    outprogram<<endl;

    outprogram<<"subject to c15: e2U + e5U + e8U <= z+1-1/522;"<<endl;
    outprogram<<"subject to c16: e3U + e6U + e9U <= z+1-1/522;"<<endl;
    outprogram<<endl;

    outprogram<<"subject to c17: e1U + e2U + e3U + e4U + e5U + e6U + e7U + e8U +e9U = 3*z+3;"<<endl;
    outprogram<<endl;

    outprogram<<"minimize statement: z;"<<endl;
    outprogram<<endl;

    for(int selection=0; selection<4; selection++)
    {
        for(int x=0; x<510; x++)
        {
            for(int y=0; y<510; y++)
            {
                string t=Rbundle[x]+Cbundle[y];
                if(Reoccurrence(t))
                    continue;
                if(!BundleRule(Rbundle[x], Cbundle[y], selection))
                    continue;
                string strU="";
                    for(int c=49;c<58;c++)
                    {
                        char car=c;
                        string stt="";
                        stt.push_back(car);
                        if(t.find(stt)!=-1)
                            ;
                        else
                            strU+=stt;
                    }
                if(selection==0||selection==1)
                {
                    outprogram<<"subject to cond"<<cond<<": ";
                    for(int i=0; i<Cbundle[y].length(); i++)
                    {
                        outprogram<<"e"<<Cbundle[y][i]<<"C"<<"+";
                    }
                    outprogram<<"0<=z+120*(1-y"<<ycond<<");"<<endl;
                    cond++;
                    ycond++;
                }
                else
                {
                    outprogram<<"subject to cond"<<cond<<": ";
                    for(int i=0; i<Rbundle[x].length(); i++)
                    {
                        outprogram<<"e"<<Rbundle[x][i]<<"R"<<"+";
                    }
                    outprogram<<"0<=z+120*(1-y"<<ycond<<");"<<endl;
                    cond++;
                    ycond++;
                }
                outprogram<<"subject to cond"<<cond<<": ";
                for(int i=0; i<strU.length(); i++)
                {
                    outprogram<<"e"<<strU[i]<<"U"<<"+";
                }
                outprogram<<"0<=z+1-1/522+120*(1-y"<<ycond<<");"<<endl;
                cond++;
                ycond++;
                outprogram<<"subject to cond"<<cond<<": ";
                int interim=ycond-2;
                outprogram<<"y"<<interim<<"+y"<<interim+1<<">=1;"<<endl<<endl;
                cond++;
            }
        }
    }
    outprogram<<"end;";
    outprogram.close();
    return 0;
}




\end{lstlisting}
\pagebreak

\end{appendix}
\end{document}